\documentclass[twocolumn,showpacs,aps,pre]{revtex4-2}
\usepackage{bm}
\usepackage{amssymb,amsmath}
\usepackage{graphicx}
\usepackage{color}
\usepackage[colorlinks=true,urlcolor=blue,citecolor=blue]{hyperref}

\begin{document}
\title{Contact process with simultaneous spatial and temporal disorder}
\author{Xuecheng Ye}
\affiliation{Department of Physics, Missouri University of Science and Technology, Rolla, MO 65409, USA}
\author{Thomas Vojta}
\affiliation{Department of Physics, Missouri University of Science and Technology, Rolla, MO 65409, USA}

\begin{abstract}
We study the absorbing-state phase transition in the one-dimensional contact process under the combined influence
of spatial and temporal random disorders. We focus on situations in which the spatial and temporal disorders decouple.
Couched in the language of epidemic spreading, this means that some spatial regions are, at all times, more favorable
than others for infections, and some time periods are more favorable than others independent of spatial location.
We employ a generalized Harris criterion to discuss the
stability of the directed percolation universality class against such disorder. We then perform large-scale Monte Carlo
simulations to analyze the critical behavior in detail. We also discuss how the Griffiths singularities that accompany
the nonequilibrium phase transition are affected by the simultaneous presence of both disorders.
\end{abstract}

\date{\today}

\maketitle

\section{Introduction}
\label{sec:intro}

Macroscopic systems far from thermal equilibrium can undergo abrupt transformations between different steady states
when their external conditions are varied. These nonequilibrium phase transitions share many features with thermodynamic
(equilibrium) phase transitions including collective behavior and large-scale fluctuations. They can be found, for example,
in interface growth, chemical reactions, granular flow, and in biological problems such as population dynamics or epidemic
spreading (for reviews, see, e.g.,
Refs.\ \cite{MarroDickman99,Hinrichsen00,Odor04,HenkelHinrichsenLuebeck_book08,Tauber_book14}).

When a nonequilibrium phase transition separates an active fluctuating steady state from an inactive absorbing state in
which fluctuations completely stop, it is called an absorbing state transition. Experimental realizations of absorbing state
transitions have been observed, for example, in turbulent liquid crystals \cite{TKCS07}, periodically driven suspensions
\cite{CCGP08,FFGP11}, bacteria colony biofilms \cite{KorolevNelson11,KXNF11}, and the dynamics of
superconducting vortices \cite{OkumaTsugawaMotohashi11}.
Janssen and Grassberger \cite{Janssen81,Grassberger82} conjectured that all continuous transitions into a single absorbing state
having a scalar order parameter and short-range interactions belong to the directed percolation (DP) universality class
\cite{GrassbergerdelaTorre79}, provided they do not feature extra symmetries, conservation laws, inhomogeneities, or disorder.

Many realistic systems undergoing absorbing state transitions feature random spatial inhomogeneities (i.e., spatial disorder)
or random variations of their external parameters with time (i.e., temporal disorder).
The question of how disorder
affects absorbing state transitions (and the DP universality class in particular) has attracted significant attention during the
last two decades or so. According to the Harris criterion \cite{Harris74}, the DP critical point
is unstable against spatial disorder because its correlation length exponent $\nu_\perp$ violates
the  inequality $d\nu_\perp>2$  in all physical dimensions, $d=1, 2$, and 3. The DP critical point is also unstable against
temporal disorder because its correlation time exponent $\nu_\parallel=z\nu_\perp$ violates Kinzel's generalization
\cite{Kinzel85} $\nu_\parallel > 2$ of the Harris criterion (see Ref.\ \cite{VojtaDickman16} for an extension
of the Harris criterion to general spatio-temporal disorder).

Spatial disorder has been demonstrated to have dramatic effects on the DP universality class.
Hooyberghs et al.\ \cite{HooyberghsIgloiVanderzande03,*HooyberghsIgloiVanderzande04}
developed a strong-disorder renormalization group (RG) \cite{MaDasguptaHu79,IgloiMonthus05} method and predicted the
transition to be governed by an unconventional infinite-randomness critical point. It is accompanied by
strong power-law Griffiths singularities \cite{Griffiths69,Noest86,*Noest88} in the parameter region close to the
transition.
The infinite-randomness critical point scenario was confirmed by large-scale Monte Carlo simulations in one, two, and
three space dimensions \cite{VojtaDickison05,OliveiraFerreira08,VojtaFarquharMast09,Vojta12}. Similar critical behavior
was also observed in diluted systems near the percolation threshold \cite{VojtaLee06,*LeeVojta09} and in systems
featuring aperiodic order \cite{BarghathiNozadzeVojta14}.

More recently, the effects of temporal disorder on the DP universality class were analyzed by means of a
real-time ``strong-noise'' RG \cite{VojtaHoyos15}. This method predicts that the disorder strength diverges with
increasing time scale at criticality, and the probability distribution of the
density becomes infinitely broad, even on a logarithmic scale. This infinite-noise critical
behavior can be understood as the temporal counterpart of infinite-randomness critical behavior
in spatially disordered systems, but with exchanged roles of space and time.
The RG predictions were later confirmed by Monte Carlo simulations \cite{BarghathiVojtaHoyos16,BarghathiTackkettVojta17}.
In addition, Vazquez et al.\ \cite{VBLM11} identified a temporal analog of the Griffiths
phase in spatially disordered systems that features an unusual power-law relation between lifetime
and system size on the active side of the phase transition.

Although the effects of pure spatial disorder and pure temporal disorder have been studied in some detail,
their simultaneous influence on absorbing state transitions has received much less attention. This is likely due
to the fact that uncorrelated spatiotemporal disorder is an irrelevant perturbation at the clean DP critical
point and thus not expected to change the critical behavior (see, e.g., Ref \cite{Hinrichsen00}).
However, many experimental applications do not lead to uncorrelated spatiotemporal disorder. Consider, for example,
an epidemic spreading in an inhomogeneous environment under conditions that fluctuate with time. If the locations
of favorable spatial regions do not change with time, and if favorable conditions in time apply uniformly to
the entire population, the resulting spatiotemporal disorder features infinite-range correlations and is thus
expected to be a relevant perturbation at the clean DP critical point.

In the present paper, we combine generalizations of the Harris criterion, optimal fluctuation arguments,
and large-scale Monte Carlo simulations to investigate the fate of the nonequilibrium phase transition in
the contact process \cite{HarrisTE74} under the influence of such spatiotemporal disorder. We find that
adding weak temporal disorder to a spatially disordered system does not change the infinite-randomness
critical behavior. Analogously, adding weak spatial disorder to a temporally disordered system does not
affect the infinite-noise critical behavior. We also explore the fate of the transitions if both disorders
are strong. In addition, we demonstrate that the functional form of the Griffiths singularities changes in
the simultaneous presence of both disorders.

Our paper is organized as follows. The contact process and our implementation of the spatiotemporal
disorder are introduced in Sec.\ \ref{sec:cp}. Section \ref{sec:scaling} briefly summarizes, what is known
about the phase transition in the clean contact process, the spatially disordered contact process,
and the temporally disordered contact process. The effects of rare regions and the resulting Griffiths
singularities in the contact process with purely spatial or purely temporal disorder are summarized in Sec.\
\ref{sec:griffiths}. The computer simulations methods are introduced in Sec.\ \ref{sec:mc}. Sections
\ref{sec:results} and \ref{sec:results-griffiths} are  devoted to our results for the contact process
in the presence of spatiotemporal disorder. We conclude in Sec.\ \ref{sec:conclusions}.

\section{Contact process}
\label{sec:cp}

The non-equilibrium phase transition in the clean contact process is well studied and belongs to the DP universality class \cite{HarrisTE74}. We consider a $d$-dimensional hypercubic lattice in which each site can be either active (infected) or inactive (healthy). As the time progresses, an active site can either infect its lattice neighbors or spontaneously become inactive. More specifically, the time evolution of the contact process is a continuous-time Markov process during which the infected sites heal at rate $\mu$, and infect their inactive neighbors at infection rate $\lambda$. Thus an inactive site becomes active at rate $\lambda n/2d$. Here, $n$ stands for the number of active neighbors. The long-time fate of the contact process is determined by the ratio between the infection rate $\lambda$ and the healing rate $\mu$. (Since only the ratio matters, $\mu$ can be set to unity without loss of generality.)

For small infection rate $\lambda$, the healing process is favored. Because of the lack of new infections, all active sites will heal eventually. The system thus ends up in the absorbing healthy state. This is called the inactive phase. For large infection rate $\lambda$, the active sites proliferate and never die out. This is called the active phase. The active and inactive phases are separated by a transition in the DP universality class.

We now introduce spatial and temporal disorder into the infection rate $\lambda$ by defining
the local infection rate $\lambda(x,t)$ at lattice site $x$ and time $t$ with a multiplicative structure,
\begin{equation}
\lambda(x,t) = \lambda_0 f(x) g(t).
\label{eq:lambda}
\end{equation}
Here, the random variables $f(x)$ and $g(t)$ are nonnegative and independent of each other. They are characterized by the averages
\begin{equation}
\langle f(x) \rangle = \bar f ~,\quad \langle g(t)\rangle =\bar g
\end{equation}
and short-range correlations
\begin{eqnarray}
\langle f(x)f(x')\rangle -\bar f^2 &=& \sigma_f^2 \delta(x-x')  ~, \\ \langle g(t)g(t')\rangle -\bar g^2 &=& \sigma_g^2 \delta(t-t')
\end{eqnarray}
The multiplicative structure implies that favorable (for the infection) spatial regions do not change with time, and favorable time intervals apply to the whole system. In other words, the disorder contains infinite-range correlations in space and time. This is reflected in the covariance function of $\lambda(x,t)$ which reads
\begin{eqnarray}
G(x-x',t-t') &=& \langle \lambda(x,t) \lambda(x',t') \rangle - \langle \lambda(x,t)\rangle \langle \lambda(x',t')\rangle \nonumber\\
             &=& ~~\lambda_0^2 \sigma_f^2\sigma_g^2  \delta(x-x') \delta(t-t') \nonumber \\
             &&  +\lambda_0^2 \sigma_f^2 \bar g^2 \delta(x-x') \nonumber \\
             &&  + \lambda_0^2 \sigma_g^2 \bar f^2 \delta(t-t')~.
\label{eq:lambda_cov}
\end{eqnarray}
Here the first term represents uncorrelated spatiotemporal disorder, the second term is perfectly correlated in time, and the last term is perfectly correlated in space. Purely spatial disorder can be understood as a special case of (\ref{eq:lambda}) with
$g=\textrm{const}$. Analogously, purely temporal disorder emerges for $f=\textrm{const}$.

\section{Scaling scenarios}
\label{sec:scaling}

In this section, we briefly summarize what is known about the critical behavior of the nonequilibrium
phase transitions in the clean contact process, the contact process with purely spatial disorder,
and the contact process with purely temporal disorder.

\subsection{Clean contact process: conventional power-law critical behavior}
\label{subsec:power-law}

The (clean) DP universality class features three independent critical exponents which can be
chosen to be $\beta$, $\nu_\perp$, and $z$ (see, e.g., Ref.\ \cite{Hinrichsen00}).
The order parameter exponent $\beta$ controls how the steady state density $\rho_{\rm stat}$
varies as  the infection rate $\lambda$ approaches its critical value $\lambda_c$ from the active side of the transition,
\begin{equation}
\rho_{\rm stat} \sim (\lambda-\lambda_c)^\beta \sim r^\beta~,
\label{eq:beta}
\end{equation}
with $r=(\lambda-\lambda_c)/\lambda_c$ the dimensionless distance from criticality.
The correlation length exponent $\nu_\perp$ controls the
divergence of the (spatial) correlation length $\xi_\perp$,
\begin{equation}
\xi_\perp \sim |r|^{-\nu_{\perp}}~,
\label{eq:nu}
\end{equation}
and the dynamical exponent $z$ relates the correlation time $\xi_\parallel$ to the correlation length,
\begin{equation}
\xi_\parallel \sim \xi_\perp^z~. \label{eq:powerlawscaling}
\end{equation}

The density $\rho$ of active sites as a function of the distance $r$ from criticality, the time $t$, and the system size $L$
fulfills the homogeneity relation
\begin{equation}
\rho(r,t,L) = \ell^{\beta/\nu_\perp} \rho(r \ell^{-1/\nu_\perp},t \ell^z, L \ell)
\label{eq:rho}
\end{equation}
where $\ell$ is an arbitrary dimensionless length scale factor.
The survival probability $P_s$ is the probability that an active cluster survives to time $t$
 if the epidemic starts at time 0 from a single infected site in an otherwise inactive lattice.
In the DP universality class, $P_s$ has the same scaling form as the density of active sites (\ref{eq:rho})
\footnote{This stems from a special time reversal symmetry \cite{GrassbergerdelaTorre79}.
   At general absorbing state transitions, e.g., with several
    absorbing states, the survival probability scales with an exponent $\beta'$ which may
    be different from $\beta$ (see, e.g., \protect{\cite{Hinrichsen00}}).},
\begin{equation}
P_s(\Delta,t,L) = \ell^{\beta/\nu_\perp} P_s(\Delta \ell^{-1/\nu_\perp},t \ell^z, L \ell)~.
\label{eq:Ps}
\end{equation}
The pair connectedness function $C(x,t)$ is given by the probability that site
$x$ is infected at time $t$ when the time evolution starts from a single infected site at
$x=0$ and time $t=0$. The scale dimension of $C$ is $2\beta/\nu_\perp$
because it involves a product of two densities
 \footnote{This relation relies on hyperscaling; it is only valid below the
    upper critical dimension $d_c^+$, which is four for directed percolation.}.
This implies the scaling form
\begin{equation}
C(r,x,t,L) = \ell^{2\beta/\nu_\perp} C(r \ell^{-1/\nu_\perp}, x \ell, t \ell^z, L \ell)~.
\label{eq:C}
\end{equation}
The number $N_s$ of sites in an active cluster growing from a single seed can be calculated
by integrating $C$ over all $x$,
\begin{equation}
N_s(r,t,L) = \ell^{2\beta/\nu_\perp - d} N_s(r \ell^{-1/\nu_\perp},t \ell^z, L \ell)~.
\label{eq:Ns}
\end{equation}
Because the mean-square radius $R$ of this active cluster has the dimension of a length,
its scaling form reads
\begin{equation}
R(r,t,L) = \ell^{-1} R(r \ell^{-1/\nu_\perp},t \ell^z, L \ell)~.
\label{eq:R}
\end{equation}

The time dependencies of $\rho$, $P_s$, $N_s$ and $R$  at the critical point
$r=0$ and in the thermodynamic limit $L\to \infty$ can be easily derived from Eqs.\
(\ref{eq:rho}) to (\ref{eq:R}) by setting the scale factor $\ell$ to suitable values.
In the long-time limit, the density of infected sites and the survival
probability are expected to follow the relations
\begin{equation}
\rho(t) \sim t^{-\delta}, \qquad P_s(t) \sim t^{-\delta}
\label{eq:Ps_t}
\end{equation}
with $\delta=\beta/(\nu_\perp z)$. The mean-square radius and the number of sites of an active
cluster starting from a single seed site behave as
\begin{equation}
R(t) \sim t^{1/z}, \qquad N_s(t) \sim t^\Theta~.
\label{eq:Ns_t}
\end{equation}
Here, $\Theta=d/z - 2\beta/(\nu_\perp z)$ is the critical initial slip exponent.
These results imply that $\Theta$, $\delta$, and $z$ are not independent, they fulfill the
hyperscaling relation $\Theta + 2 \delta= d/z$.

Highly accurate estimates of the critical exponents for the clean DP universality class in $d=1$ dimensions
were computed by series expansions \cite{Jensen99}: $\beta=0.276486$, $\nu_\perp=1.096854$, $z=1.580745$,
$\delta=0.159464$, and $\Theta=0.313686$.

The clean correlation length exponent violates Harris' inequality $d\nu_\perp > 2$ \cite{Harris74}. Analogously,
the exponent combination $\nu_\parallel=z\nu_\perp$ violates the corresponding inequality $\nu_\parallel > 2$
for temporal disorder \cite{Kinzel85}. Consequently, the clean DP critical behavior is unstable against both
purely spatial disorder and purely temporal disorder.

\subsection{Spatially disordered contact process: infinite-randomness critical behavior}
\label{subsec:infinite-randomness}

Hooyberghs et al.\ \cite{HooyberghsIgloiVanderzande03,*HooyberghsIgloiVanderzande04}
employed a strong-disorder renormalization group (RG) \cite{MaDasguptaHu79,IgloiMonthus05} method to demonstrate that
the nonequilibrium phase transition in the spatially disordered contact process is governed by an exotic infinite-randomness
critical point in the same universality class as the random transverse-field Ising model \cite{Fisher92,*Fisher95}.
This was later verified by Monte-Carlo simulations in one, two, and
three space dimensions \cite{VojtaDickison05,OliveiraFerreira08,VojtaFarquharMast09,Vojta12}.

A key difference between a conventional critical point and an infinite-randomness critical point
is the replacement of the power-law relation (\ref{eq:powerlawscaling}) between correlation length and time
by an exponential (activated) one,
\begin{equation}
\ln(\xi_\parallel/t_0) \sim \xi_\perp^\psi~. \label{eq:activatedscaling}
\end{equation}
Here $\psi$ is the so-called tunneling exponent, and $t_0$ is a microscopic time scale.
This exponential relation implies that the dynamical exponent $z$ is formally infinite at an infinite-randomness
critical point. In contrast, the static scaling relations
remain of power-law type, i.e., eqs.\ (\ref{eq:beta}) and (\ref{eq:nu}) remain valid.

The scaling forms of disorder-averaged observables can be obtained by
simply substituting the variable combination $\ln(t/t_0) \ell^\psi$  for $t \ell^z$
in the arguments of the scaling functions, yielding
\begin{equation}
\rho(r,\ln(t/t_0),L) = \ell^{\beta/\nu_\perp} \rho(r \ell^{-1/\nu_\perp},\ln(t/t_0) \ell^\psi, L \ell)~, \label{eq:rho_activated}
\end{equation}
\begin{equation}
P_s(r,\ln(t/t_0),L) = \ell^{\beta/\nu_\perp} P_s(r \ell^{-1/\nu_\perp},\ln(t/t_0) \ell^\psi,L \ell)~, \label{eq:Ps_activated}
\end{equation}
\begin{equation}
 N_s(r,\ln(t/t_0),L) = \ell^{2\beta/\nu_\perp -d} N_s(r
\ell^{-1/\nu_\perp},\ln(t/t_0) \ell^\psi,L \ell)~, \label{eq:Ns_activated}
\end{equation}
\begin{equation}
 R(r,\ln(t/t_0),L) = \ell^{-1} R(r \ell^{-1/\nu_\perp},\ln(t/t_0) \ell^\psi,L \ell)~.  \label{eq:R_activated}
\end{equation}

The resulting critical time dependencies of  $\rho$, $P_s$, $N_s$, and $R$
are logarithmic (in the thermodynamic limit),
\begin{eqnarray}
\rho(t) \sim [\ln(t/t_0)]^{-\bar\delta}, \qquad P_s(t) \sim [\ln(t/t_0)]^{-\bar\delta}~,
\label{eq:Ps_t_activated} \\
R(t) \sim [\ln(t/t_0)]^{1/\psi}, \qquad N_s(t) \sim [\ln(t/t_0)]^{\bar\Theta}~,
\label{eq:Ns_t_activated}
\end{eqnarray}
with $\bar\delta=\beta/(\nu_\perp \psi)$ and $\bar\Theta=d/\psi-2\beta/(\nu_\perp \psi)$.

Within the strong-disorder renormalization group approach, the
critical exponents of the spatially disordered one-dimensional contact process can be calculated
exactly. Their numerical values are $\beta=0.38197$, $\nu_\perp=2$, $\psi=0.5$,
$\bar\delta=0.38197$, and $\bar\Theta=1.2360$.

\subsection{Temporally disordered contact process: infinite-noise critical behavior}
\label{subsec:infinite-noise}

To attack the problem of temporal disorder in the contact process, Vojta and Hoyos \cite{VojtaHoyos15}
developed a real-time strong-noise renormalization group that can be understood as the temporal
counterpart of the strong-disorder renormalization group for spatially disordered systems.
This renormalization group predicts (in any finite dimensionality $d$) a Kosterlitz-Thouless
\cite{KosterlitzThouless73} type transition at which the critical fixed point is the end
point of a line of fixed points that describe the ordered phase.
Consequently, observables at criticality show the same qualitative behavior as in the active phase,
except for logarithmic corrections. This can be expressed in the following heuristic scaling theory
\cite{BarghathiVojtaHoyos16}.

The density of active sites fulfills the scaling form
\begin{equation}
\rho (r,t,L) = (\ln \ell)^{-\beta/\bar\nu_\perp} \rho(r (\ln \ell)^{1/\bar\nu_\perp}, t\ell^{-z}, L\ell^{-1})
\label{eq:rho_scaling_temporal}
\end{equation}
with order parameter exponent $\beta=1/2$, correlation length exponent $\bar\nu_\perp=1/2$, and
dynamical exponent $z=1$. The scaling combination $r (\ln \ell)^{1/\bar\nu_\perp}$ reflects the exponenial dependence of the correlation length $\xi_\perp$ on the distance $r$ from criticality. Because the time reversal symmetry of DP \cite{GrassbergerdelaTorre79} is still
valid in the presence of temporal disorder, the survival probability has the same scaling form,
\begin{equation}
P_s (r,t,L) = (\ln \ell)^{-\beta/\bar\nu_\perp} P_s(r (\ln \ell)^{1/\bar\nu_\perp}, t \ell^{-z}, L\ell^{-1})~.
\label{eq:Ps_scaling_temporal}
\end{equation}
The cloud of active sites originating from a single infected seed site spreads ballistically, apart from
logarithmic corrections, yielding the scaling forms
\begin{eqnarray}
N_s (r,t,L) &=& \ell^d(\ln \ell)^{-y_N} N_s(r (\ln \ell)^{1/\bar\nu_\perp}, t\ell^{-z}, L\ell^{-1})~,~
\label{eq:Ns_scaling_temporal}\\
R (r,t,L) &=& \ell(\ln \ell)^{-y_R} R(r (\ln \ell)^{1/\bar\nu_\perp}, t\ell^{-z}, L\ell^{-1})~.
\label{eq:R_scaling_temporal}
\end{eqnarray}
The exponents $y_N$ and $y_R$ that govern the logarithmic corrections are not independent of
each other. Because $N_s \sim P_s \rho R^d$, they must fulfill the relation $y_N = 2\beta/\bar\nu_\perp  + d y_R$.

Setting $L=\infty$, $r=0$, and $\ell=t^{1/z}=t$ in the scaling forms (\ref{eq:rho_scaling_temporal}) to (\ref{eq:R_scaling_temporal})
gives the time dependencies of the observables at criticality,
\begin{eqnarray}
\rho(t) &\sim& (\ln t)^{-\bar\delta}~, \quad\qquad P_s(t) \sim (\ln t)^{-\bar\delta} \label{eq:rho_t_temporal}\\
R(t) &\sim& t^{1/z} (\ln t)^{-y_R}~,  \quad N_s(t) \sim t^\Theta (\ln t)^{-y_N} \label{eq:Ns_t_temporal}
\end{eqnarray}
with $\bar\delta=\beta/\bar \nu_\parallel= 1$ and $\Theta=d/z = d$.

This scaling theory was confirmed by large-scale Monte Carlo simulations of the contact process with temporal disorder
in one and two space dimensions \cite{BarghathiVojtaHoyos16}. The simulations resulted in the estimates
$y_N=3.6(4)$ and $y_R=1.7(3)$ for the exponents governing the logarithmic corrections in one dimension
\footnote{The numbers in brackets indicate the errors of the last digits.}.

\section{Rare events and Griffiths singularities}
\label{sec:griffiths}

Spatial and temporal disorder do not only destabilize the DP critical behavior, rare strong disorder fluctuations
also lead to unusual singularities, the Griffiths singularities \cite{Griffiths69,Vojta06}
in an entire parameter region around the transition.
This section briefly summarizes the rare region effects in the contact process with purely spatial disorder,
and in the contact process with purely temporal disorder.

\subsection{Spatial disorder}
\label{subsec:griffiths-spatial}

The inactive phase of a spatially disordered contact process can generally be divided into two regions.
Far away from criticality (i.e., for sufficiently small infection rate), the system approaches the
absorbing state exponentially fast in time, just as in the absence of disorder. This is the conventional
inactive phase. For infection rates closer to the disordered critical point, the system may feature
large spatial regions that are locally in the active phase even though the system as a whole is still inactive.
Because these regions are of finite size, they cannot support a nonzero steady-state density, but
their density decay is very slow since it requires a rare density fluctuation of the entire region
\cite{Noest86,*Noest88}. The range in parameter space for which such rare locally active spatial regions exist is
called the (inactive) Griffith phase.

The contribution $\rho_{RR}(t)$ of the rare regions to system's density can be easily estimated as
\begin{equation}
\rho_{RR}(t) \sim \int dL_{RR} ~L_{RR}^d ~w(L_{RR}) \exp\left[-t/\tau(L_{RR})\right]~,
\label{eq:rho_RR}
\end{equation}
where $w(L_{RR})$ is the probability for finding a rare region of linear size $L_{RR}$,
and $\tau(L_{RR})$ is its decay time. For uncorrelated or short-range correlated disorder,
the rare region probability is given by $w(L_{RR}) \sim \exp(- b L_{RR}^d)$ (up to pre-exponential
factors). The decay time reads $\tau(L_{RR}) \sim \exp(a L_{RR}^d)$ because a coordinated fluctuation
of the entire rare region is required to take it to the absorbing state.

In the long-time limit, the integral (\ref{eq:rho_RR}) can be evaluated using the saddle point method,
yielding an anomalous power-law decay of the density in the Griffiths phase,
\begin{equation}
\rho(t) \sim  t^{-b/a}  = t^{-d/z'}~,
\label{eq:griffithspower}
\end{equation}
rather then the exponential decay in the conventional inactive phase.
Here $z'=da/b$ is the \emph{nonuniversal} Griffiths dynamical exponent. The survival probability $P_s$ shows
exactly the same time dependence.
The behavior of $z'$ close to the infinite-randomness critical point $\lambda_c$ follows from the strong-disorder
renormalization group \cite{HooyberghsIgloiVanderzande04,Fisher95},
\begin{equation}
z' \sim  |\lambda-\lambda_c|^{-\psi\nu_\perp}~,
\label{eq:z_prime}
\end{equation}
where $\psi$ and $\nu_\perp$ are the critical exponents of the infinite-randomness critical point.
Similar rare region effects also exist in the active phase where they govern the approach
to the nonzero steady-state density.

\subsection{Temporal disorder}
\label{subsec:griffiths-temporal}

The temporal Griffiths phase, introduced by Vazquez et al.\ \cite{VBLM11},
is the part of the active phase in which the life time $\tau_L$
of a finite-size sample shows an anomalous (non-exponential) dependence on the system size
$L$.

The temporal Griffiths behavior is the result of rare, long time intervals during
which the system is temporarily on the inactive side of the transition. The probability
of finding such a time interval of length $T_{RR}$ depends exponentially on its length,
$w(T_{RR}) \sim \exp(-b T_{RR})$ (neglecting pre-exponential factors). During $T_{RR}$,
the density of active sites decays exponentially as $\rho \sim \exp(-a t)$. Because
the typical life time of a system of linear size $L$ can be estimated as time when the
density reaches the value $L^{-d}$, a system of size $L$ will die during a rare time
interval of length $T_{RR} \sim (d/a) \ln L$. The characteristic time it takes for
such a rare time interval to appear is given by $\tau \sim w^{-1}(T_{RR})\sim  \exp (b T_{RR})$.
Consequently, the life time $\tau$ of a finite-size system in the temporal Griffiths phase shows
a power-law dependence on its size $L$,
\begin{equation}
\tau(L) \sim L^{db/a} = L^{d/\kappa}~.
\label{eq:griffithspower_temporal}
\end{equation}

The infinite-noise renormalization group \cite{VojtaHoyos15,BarghathiVojtaHoyos16} predicts that
the Griffiths exponent $\kappa= a/b$ take the value $\kappa_c =d$ right at criticality. $\kappa$
decreases with increasing distance from criticality and is expected to vanish
at the boundary between the temporal Griffiths phase and the conventional active phase (in which
the life time increases exponentially with system size).
The temporal Griffiths behavior has been confirmed by Monte Carlo simulations of the contact
process with temporal disorder in one and two space dimensions \cite{BarghathiVojtaHoyos16}.

\section{Simulations methods}
\label{sec:mc}

Our computer simulations focus on the case of one space dimension. The numerical implementation of the one-dimension contact process follows the method developed by Dickman \cite{Dickman99}. We start at $t$ = 0 from a system with at least one active site. For each time step, we follow this sequence: First, an active site is randomly chosen from all $N_a$ active sites. Then we randomly let this site infect one of its neighbors with probability $\lambda(x,t)/[\lambda(x,t)+1]$ or become inactive with probability $1/[\lambda(x,t)+1]$. If the infection process is chosen, only a single neighbor is infected, chosen randomly. The time increment associated with this sequence is $1/N_a$.

As discussed in Sec.\ \ref{sec:cp}, the local infection rates take the form $\lambda(x,t) = \lambda_0 f(x) g(t)$,
where $\lambda_0$ is the control parameter used to tune the phase transition, and $f(x)$ and $g(t)$ are independent
random variables. (In the following, we will drop the subscript $0$ from $\lambda_0$ if the meaning is clear.)
For the computer simulations, we employ the binary probability distribution
\begin{equation}
P(f)=(1-p)\delta(f-1)+p(f-c)~,
\label{eq:ax}
\end{equation}
with $0 < c \leq 1$. This means the local infection rate is reduced by a factor $c$ with probability $p$.
$g(t)$ is piecewise constant over short time intervals of length $\Delta t =6$, i.e., $g(t) = g_n$ for $t_{n+1} > t > t_n$ with $t_n= n \Delta t$. The $g_n$ follow a binary probability distribution
\begin{equation}
P(g_n)=(1-p_t)\delta(g_n-1)+p_t(g_n-c_t)~.
\label{eq:bt}
\end{equation}

We study two sequences of parameters. The first sequence starts from (strong) purely spatial disorder, adding an increasing amount of temporal disorders ($p = 0.3$, $c = 0.2$, $p_t = 0.2$ and $c_t$ varying from 1.0 to 0.12). The other sequence starts from (strong) purely temporal disorder and adds an increasing amount of spatial disorder ($p_t = 0.2$, $c_t = 0.05$, $p = 0.2$ and $c$ varying from 1.0 to 0.05).

For each parameter set $\lambda_0$,  $p$, $c$, $p_t$ and $c_t$, the results are averaged over many disorder realizations (between $700$  and $5\times 10^{5}$). We employ two types of simulation runs, (i) decay simulations in which the system starts with all sites being active. In this case, we perform one simulation run per disorder configuration and observe the active site density $\rho (t)$. (ii) Spreading simulation start with a single active seed site only. In this case, we perform 5 to $10^5$ runs per disorder configuration and analyze the survival probability $P_s(t)$, the average number of active sites $N_s(t)$ and the (mean-square) radius $R(t)$ of the active cloud. In order to eliminate the finite-size effects for spreading runs, the system size is chosen to be much larger
than the maximum active cloud size.

\section{Results: Critical behavior}
\label{sec:results}
\subsection{Generalized Harris criterion}
\label{subsec:results-overview}

The Harris criterion $d\nu_\perp >2$ controls the stability of a clean critical point against uncorrelated (or short-range correlated) purely spatial disorder. Analogously, the inequality $\nu_\parallel > 2$ governs the stability against uncorrelated purely temporal
disorder \cite{Kinzel85}. As pointed out in Sec.\ \ref{subsec:power-law}, the clean DP critical point is unstable against both
purely spatial disorder and purely temporal disorder because its critical exponents violate both inequalities.

The effects of general spatiotemporal disorder can be ascertained by means of the generalized Harris criterion \cite{VojtaDickman16}. It predicts that a critical point is (perturbatively) stable against weak spatiotemporal disorder,
if the disorder covariance function $G(x,t)$ fulfills the condition
\begin{equation}
\xi_\perp^{2/\nu_\perp -d}\xi_\parallel^{-1} \int_{-\xi_\perp/2}^{\xi_\perp/2} d^dx \int_{-\xi_\parallel/2}^{\xi_\parallel/2}dt \,  G(x,t) \to 0
\label{eq:generalized_criterion}
\end{equation}
as the critical point is approached, i.e, for $\xi_\perp, \xi_\parallel \to \infty$ with the appropriate scaling relation
between $\xi_\perp$ and $\xi_\parallel$. For power-law dynamical scaling this means $\xi_\parallel \sim \xi_\perp^z$, and for activated scaling $\ln(\xi_\parallel/t_0) \sim \xi_\perp^\psi$.

For completely uncorrelated spatiotemporal disorder with $G(x,t) \sim \delta(x)\delta(t)$, the l.h.s.\ of Eq.\ (\ref{eq:generalized_criterion}) behaves as $\xi_\perp^{2/\nu_\perp -d}\xi_\parallel^{-1}$. The resulting stability criterion thus reads $(d+z)\nu_\perp >2$ in the case of power law dynamical scaling. The clean DP critical exponents fulfill this inequality implying that uncorrelated spatiotemporal disorder is not a relevant perturbation, as was already pointed out in the literature (see, e.g.,
Ref.\ \cite{Hinrichsen00}).

Let us now apply the generalized Harris criterion to the disorder (\ref{eq:lambda}) studied in this paper. Inserting the covariance function (\ref{eq:lambda_cov}), $G(x,t) = \lambda_0^2 \sigma_f^2\sigma_g^2  \delta(x) \delta(t) +\lambda_0^2 \sigma_f^2 \bar g^2 \delta(x)  + \lambda_0^2 \sigma_g^2 \bar f^2 \delta(t)$, into Eq.\ (\ref{eq:generalized_criterion}) produces three contributions. The first term (which represents uncorrelated disorder) goes to zero in the critical limit $\xi_\perp \to \infty$ provided the critical exponents fulfill the inequality $(d+z)\nu_\perp >2$. The second term vanishes for $d\nu_\perp >2$, and the third term vanishes for $z\nu_\perp >2$. Because the DP critical exponents violate the latter two inequalities, the disorder (\ref{eq:lambda}) is a relevant perturbation at the clean DP critical point and expected to modify the critical behavior.

The generalized Harris criterion can also be used to analyze the addition of weak temporal disorder to the already spatially disordered contact process. For purely temporal disorder, $G(x,t) \sim \delta(t)$. The l.h.s.\ of (\ref{eq:generalized_criterion}) thus behaves as $\xi_\perp^{2/\nu_\perp} \xi_\parallel^{-1}$. Because the correlation time $\xi_\parallel$ depends exponentially on the correlation length $\xi_\perp$ at the infinite-randomness critical point of the spatially disordered contact process (See Eq.\ (\ref{eq:activatedscaling})), $\xi_\perp^{2/\nu_\perp} \xi_\parallel^{-1}$ vanishes as criticality is approached,  $\xi_\perp \to \infty$. Thus, the infinite-randomness critical point is expected to be stable against weak temporal disorder. The same result also follows from Kinzel's inequality $z\nu_\perp >2$ because $z$ is formally infinite at the infinite-randomness critical point.

To study the stability of the infinite-noise critical point of the temporally disordered contact process against weak spatial disorder, we insert $G(x,t) \sim \delta(x)$ into Eq.\ (\ref{eq:generalized_criterion}). The l.h.s.\ then takes the form $\xi_\perp^{2/\nu_\perp -d}$ leading to the usual Harris inequality $d\nu_\perp >2$. As the infinite-noise critical point features Kosterlitz-Thouless critical behavior with $\ln \xi_\perp \sim |r|^{-1/2}$, the exponent $\nu_\perp$ is formally infinite. This implies that weak spatial disorder is not a relevant perturbation at the infinite-noise critical point.

The generalized Harris criterion thus predicts that adding weak temporal disorder does not modify the critical behavior of the spatially disordered contact process and vice versa. This raises the interesting question of what happens if both disorders are of comparable strength. We will return to this question in Sec.\ \ref{subsec:results-bothstrong}.

\subsection{Adding weak spatial disorder to the temporal disordered contact process}
\label{subsec:results-critical-weakspatial}

After the discussion of the generalized Harris criterion, we turn to computer simulation results. We start by adding weak spatial disorder to an already temporally disordered contact process. To this end, we simulate a sequence of systems with fixed strong temporal disorder, $p_t = 0.2$, $c_t = 0.05$ and $\Delta t =6$ and increasing spatial disorder, $p=0.2$, $c$ varying from 1.0 to 0.05. The case of purely temporal disorder  ($c = 1.0$) corresponds to the parameters studied in detail in Ref.\ \cite{BarghathiVojtaHoyos16}. Based on the generalized Harris criterion, we anticipate that the critical behavior for sufficiently weak spatial disorder remains identical to the pure temporal disorder case, albeit with a shift of the critical infection rate $\lambda_c$.

\begin{figure}
\includegraphics[width=\columnwidth]{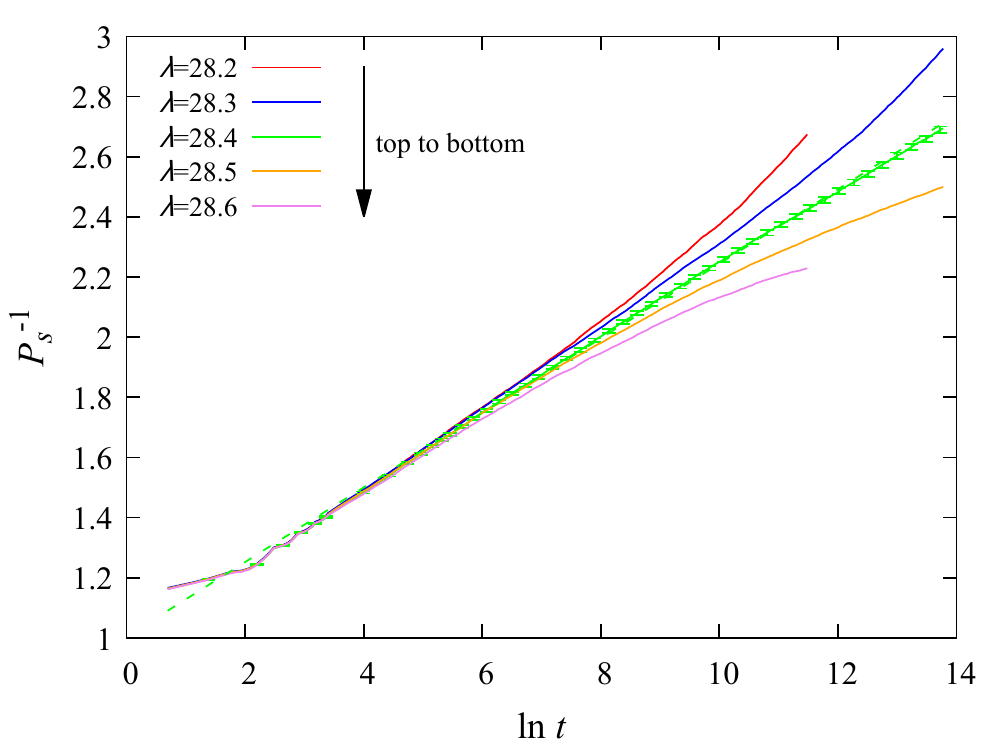}
\caption{Inverse survival probability $1/P_s$ vs ln $t$ close to criticality. The data are averages over 10000 to 20000 disorder
configurations, with 5 runs per configuration ($p_t = 0.2$, $c_t = 0.05$, $\Delta t =6$, $p = 0.2$ and $c = 0.8$). The statistical errors  of every fifth data
point of the critical curve are shown. The dashed line is a linear fit of the data from ln $t$ = 4.9 to ln $t$ = 13.8 (reduced $\chi^2 \approx 0.9$)}
\label{fig:psws}
\end{figure}

We therefore analyze the simulation data based on Eqs.\ (\ref{eq:rho_t_temporal}) and (\ref{eq:Ns_t_temporal}). Figure \ref{fig:psws} presents the inverse survival probability $1/P_s$ of spreading runs as a function of $\ln t$ for the weakest nonzero spatial disorder ($c = 0.8$).
The figure shows that the data for $\lambda = 28.4$ follow the predicted logarithmic behavior (\ref{eq:rho_t_temporal}) over almost five orders in magnitude in $t$. The data points with higher or lower $\lambda$ curve away from the straight line as expected. We therefore identify $\lambda_c = 28.4$ as the critical value for $c = 0.8$. For comparison, the critical value for the case of purely temporal disorder is $\lambda_c = 27.27$ \cite{BarghathiVojtaHoyos16}. Figure \ref{fig:psws} thus provides evidence that adding weak spatial disorder does not change the strong-noise critical behavior of the purely temporally disordered system.

\begin{figure}
\includegraphics[width=\columnwidth]{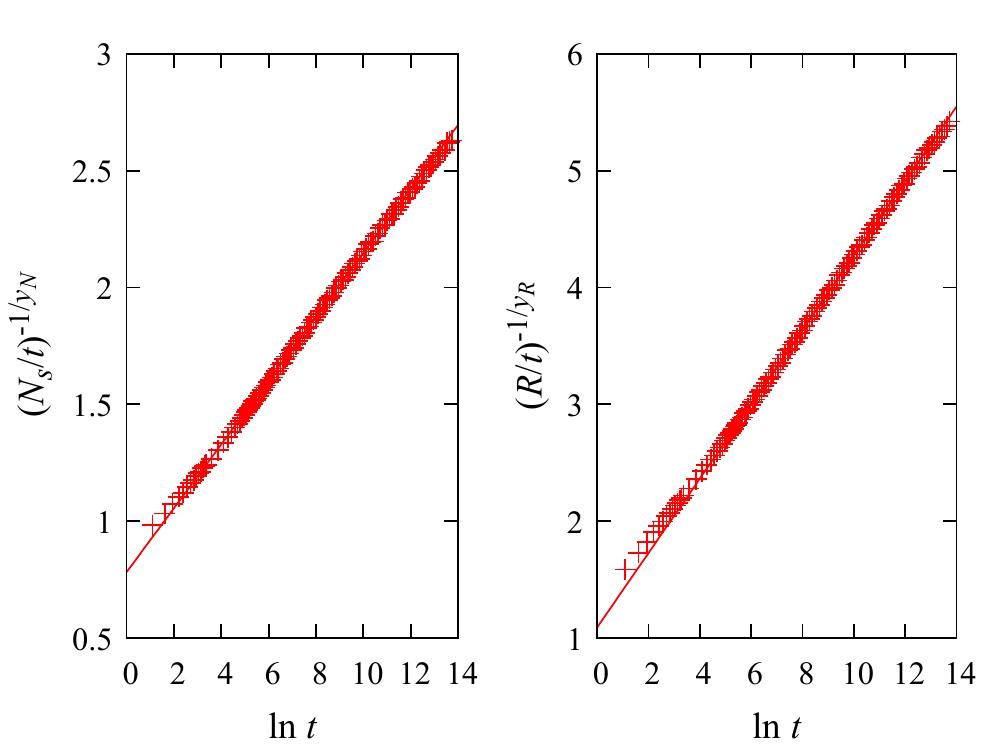}
\caption{$(N_s/t)^{-1/y_N}$ and $(R/t)^{-1/y_R}$ vs ln $t$ at criticality, $\lambda_c = 28.4$  for $p_t = 0.2$, $c_t = 0.05$, $\Delta t =6$, $p = 0.2$ and $c = 0.8$. The data are averages over 20000 disorder configurations with 5 runs for each. The exponents $y_N = 3.6$ and $y_R = 1.7$ are fixed at the values found for purely temporal disorder \cite{BarghathiVojtaHoyos16}.  The straight lines are fits of the data from ln $t$ = 6.5 to ln $t$ =13.8 with Eqs.\ (\ref{eq:Ns_t_temporal}).}
\label{fig:nsws}
\end{figure}

To further confirm this, we test Eqs.\ (\ref{eq:Ns_t_temporal}) by analyzing the number of active sites $N_s$ and the cloud radius $R$ at criticality as functions of time in Fig.\ \ref{fig:nsws}.
To make the logarithmic corrections visible, we modify $N_s$ and $R$ by dividing out the leading term $t$. We then plot $(N_s/t)^{-1/y_N}$ and $(R/t)^{-1/y_R}$ vs ln $t$, using the exponents $y_N = 3.6$ and $y_R = 1.7$ found for the case of purely temporal disorder \cite{BarghathiVojtaHoyos16}. The data follow straight lines, confirming that Eqs.\ (\ref{eq:Ns_t_temporal}) are also fulfilled.

\begin{figure}
\includegraphics[width=\columnwidth]{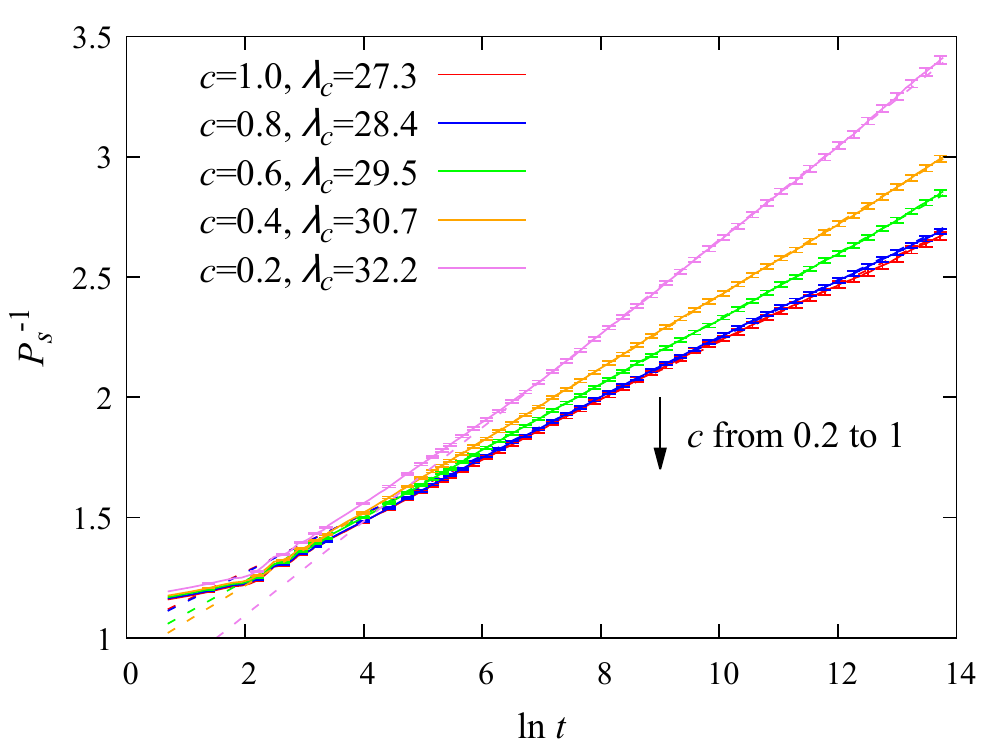}
\caption{Inverse survival probability $1/P_s$ vs ln $t$ at  criticality for different $c$ and $p_t = 0.2$, $c_t = 0.05$, $\Delta t =6$, $p = 0.2$. The data are averages over 20000 disorder configurations with 5 runs for each. The statistical errors  of every fifth data point is shown. The dashed lines are linear fits of the data.}
\label{fig:all_c}
\end{figure}

\begin{figure}
\includegraphics[width=\columnwidth]{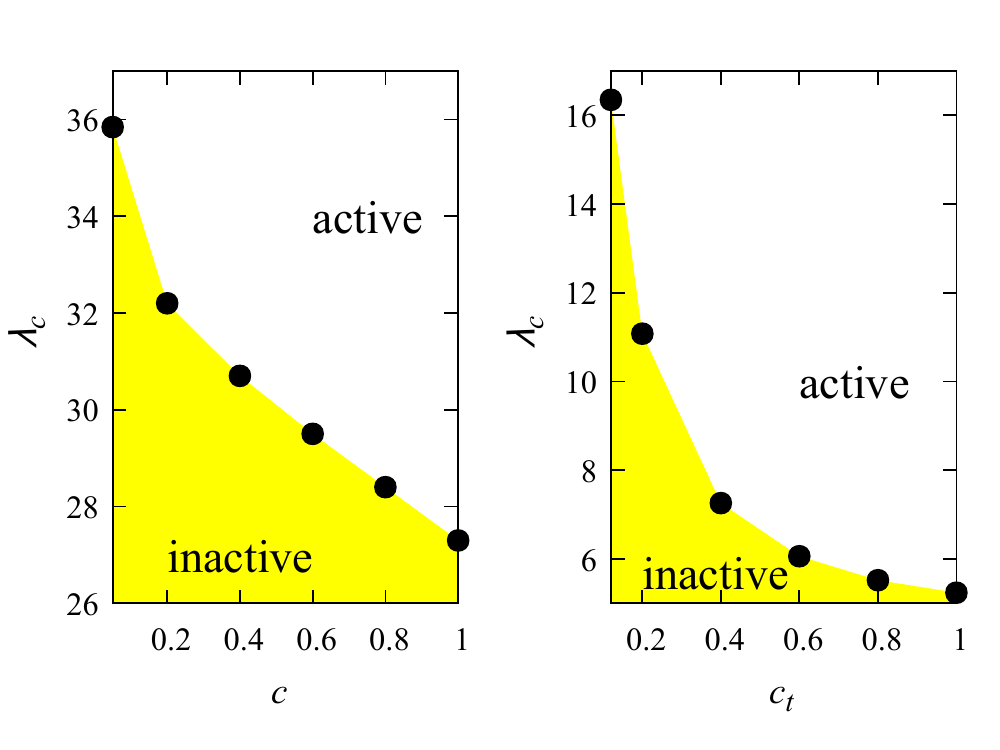}
\caption{Left: Critical infection rate $\lambda_c$ as a function of spatial disorder strength $c$ for $p_t = 0.2$, $c_t = 0.05$, $p = 0.2$. Right: Critical $\lambda_c$ as a function of temporal disorder strength $c_t$ for $p = 0.3$, $c = 0.2$, $p_t = 0.2$.}
\label{fig:pd}
\end{figure}

Now, we extend the simulations to stronger spatial disorder (decreasing $c$ towards 0). For $c = 0.6, 0.4$ and $0.2$, the critical behavior can be fitted well with the infinite-noise functional forms Eqs.\ (\ref{eq:rho_t_temporal}) and (\ref{eq:Ns_t_temporal}). This can be seem in Fig.\ \ref{fig:all_c} that shows the inverse survival probability as a function of ln $t$ of the critical curves for $c = 1, 0.8, 0.6, 0.4$ and $0.2$.
All data follow straight lines for more than three orders of magnitude in $t$, confirming Eq.\ (\ref{eq:rho_t_temporal}).
The resulting values for $\lambda_c$ are presented in Fig. \ref{fig:pd}.
When the spatial disorder is further increased, the critical behavior deviates from the infinite-noise critical behavior (\ref{eq:rho_t_temporal}) and (\ref{eq:Ns_t_temporal}). We will discuss this case in Sec. \ref{subsec:results-bothstrong}.

\subsection{Adding weak temporal disorder to spatial disorder case}
\label{subsec:results-critical-weaktemporal}

We now simulate a sequence of systems with fixed strong spatial disorder $p = 0.3$, $c = 0.2$, to which we add increasing temporal disorder with $p_t = 0.2$ and $c_t$ varying from 1.0 to 0.12. The starting point of this sequence, the purely spatially disordered system with $c_t = 1$, corresponds to the parameters studied in Ref.\ \cite{VojtaDickison05}.

\begin{figure}
\includegraphics[width=\columnwidth]{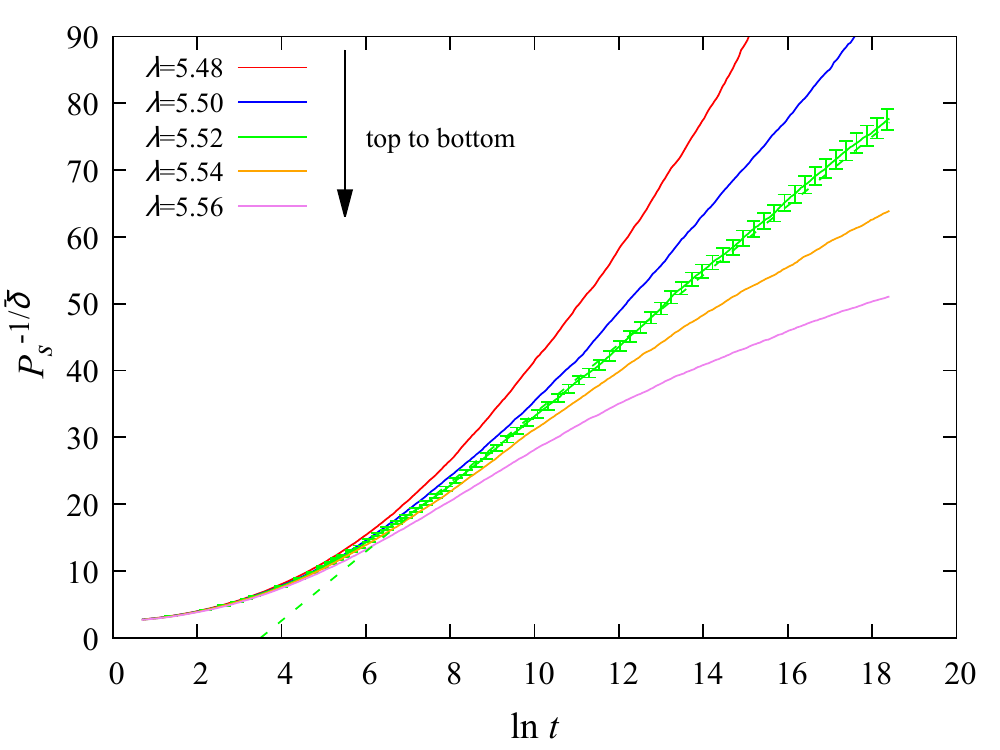}
\caption{Survival probability vs time plotted as $P_s^{-1/\bar\delta}$ vs ln $t$ close to criticality, where $\bar\delta = 0.38197$ ($p_t = 0.2$, $c_t = 0.8$, $\Delta t=6$, $p = 0.3$ and $c = 0.2$). The data are averages over 700 disorder configurations with 100 runs per configuration. The  statistical errors  of every fifth data point of the critical curve are marked. The dashed line is a linear fit of the data for $\ln t$ = 6.5 to  18.4 (reduced $\chi^2 \approx 0.9$)}
\label{fig:pswt}
\end{figure}

\begin{figure}
\includegraphics[width=\columnwidth]{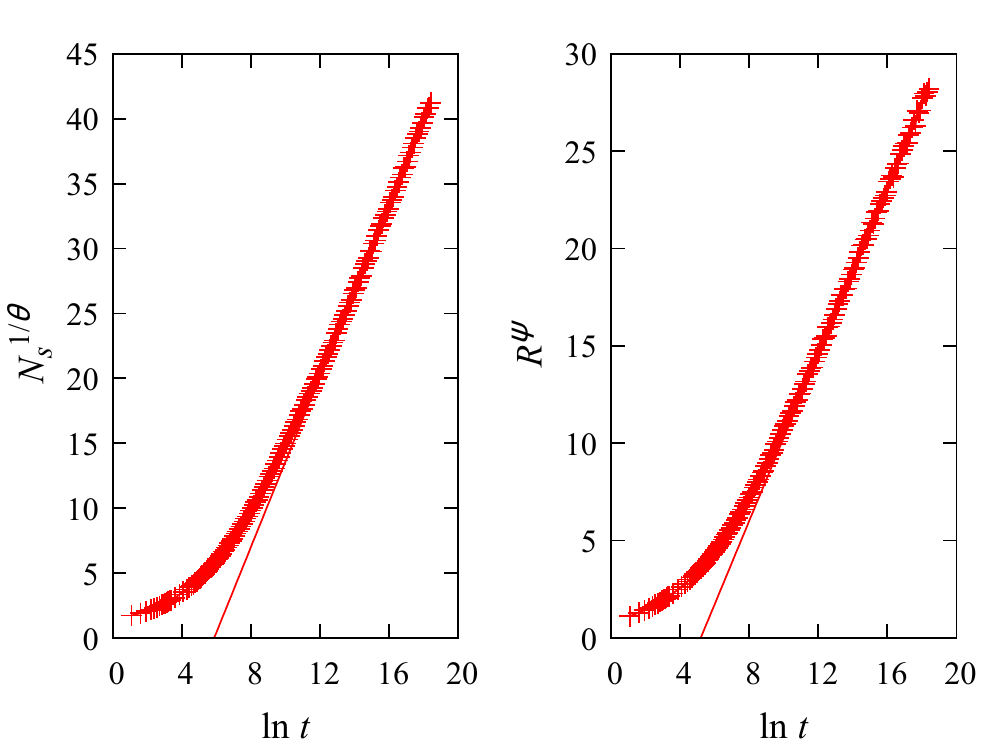}
\caption{$(N_s)^{1/\Theta}$ and $(R)^{\psi}$ vs ln $t$ at criticality $\lambda_c =5.52$ for $p_t = 0.2$, $c_t = 0.8$, $\Delta t =6$, $p = 0.3$, and $c = 0.2$. The data are averages over 700 disorder configurations with 100 runs per configuration. The values of the initial slip exponent $\Theta$ and the tunneling exponent $\psi$ are fixed at the values of the infinite randomness critical point, $\Theta = 1.2360$, $\psi = 0.5$. The solid lines represents fits to Eqs.\ (\ref{eq:Ns_t_activated}) from $\ln t = 12.7$ to 18.4.}
\label{fig:nswt}
\end{figure}

For weak temporal disorder, $c_t = 0.8$,  we anticipate the system to show the infinite-randomness critical behavior discussed in Sec. \ref{subsec:infinite-randomness}.
This is tested in Figs. \ref{fig:pswt} and \ref{fig:nswt} which present the results of spreading simulations.
Fig. \ref{fig:pswt} shows a plot of $P_s^{-1/\delta}$ vs ln $t$. The predicted critical behavior (\ref{eq:Ps_t_activated}) corresponds to a straight line in this plot. The figure demonstrates that the data for $\lambda = 5.52$ follow (\ref{eq:Ps_t_activated}) for almost five order of magnitude in $t$. This yields evidence for the infinite-randomness critical behavior. Similarly, Fig.\ \ref{fig:nswt} shows that the number of active sites sites $N_s$ and the cloud radius $R$ fulfill Eqs.\ (\ref{eq:Ns_t_activated}) for almost four orders of magnitude in $t$. We conclude that  the system is still controlled by infinite-randomness critical behavior.

\begin{figure}
\includegraphics[width=\columnwidth]{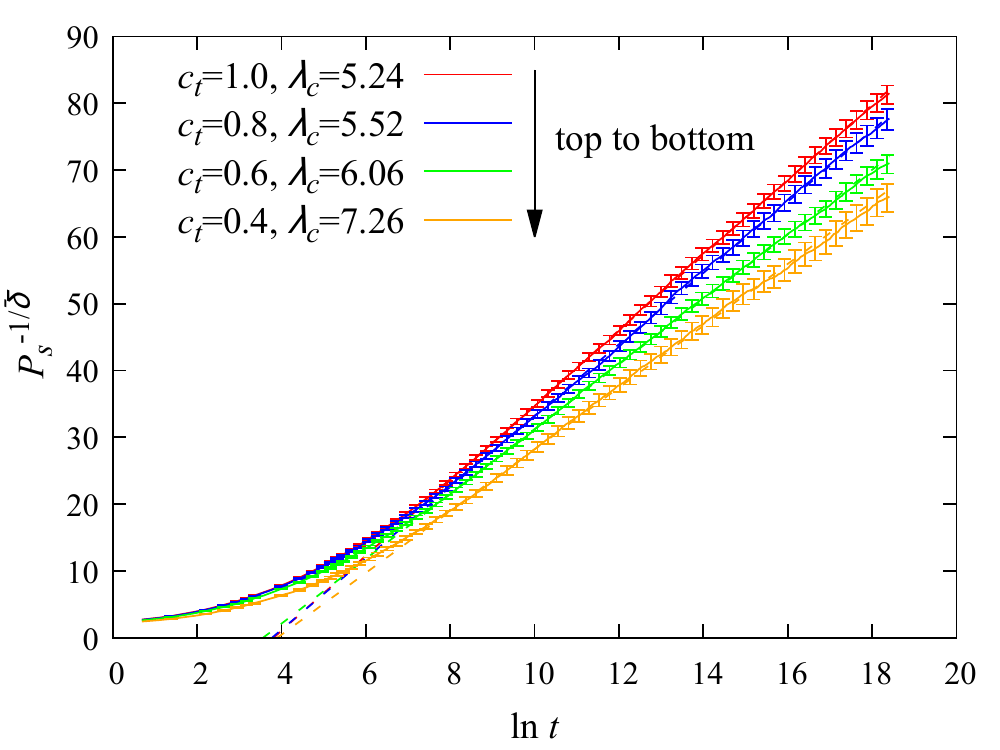}
\caption{Survival probability vs time plotted as $P_s^{-1/\bar\delta}$ vs ln $t$ at criticality for different $c_t$, where $\bar\delta = 0.38197$ ($p_t = 0.2$, $\Delta t =6$, $p = 0.3$ and $c = 0.2$). The data are averages over 700 to 1000 disorder configurations with
30 to 100 runs for each. The  statistical errors of every fifth data point are marked. The dashed lines are linear fits of the data.}
\label{fig:all_ct}
\end{figure}

We repeat this analysis for systems with stronger temporal disorder. For $c_t = 0.6$ and 0.4, we find that the critical behavior can be fitted well with the infinite-randomness expressions (\ref{eq:Ps_t_activated}) and (\ref{eq:Ns_t_activated}). This can be seen in
Fig.\ \ref{fig:all_ct}, which shows $P_s^{-1/\delta}$ vs $\ln t$ at criticality for $c_t = 1, 0.8, 0.6, 0.4$.
The data feature straight-line behavior for more than four orders of magnitude in $t$, confirming (\ref{eq:Ns_t_activated}). The critical infection rates $\lambda_c$ resulting from these simulations are shown in the phase diagram in Fig.\ \ref{fig:pd}.

For even stronger temporal disorder, the critical behavior deviates from the infinite-randomness criticality of Sec. \ref{subsec:infinite-randomness}, as will be discussed in the next section.

\subsection{Spatial and temporal disorder of comparable strength}
\label{subsec:results-bothstrong}

In Sec.\ \ref{subsec:results-critical-weakspatial}, we have demonstrated that the infinite-noise critical point of the temporally disordered contact process is stable against the addition of weak spatial disorder. Analogously, the infinite-randomness critical point of the spatially disordered contact process is stable against the addition of weak temporal disorder, as shown in Sec.\ \ref{subsec:results-critical-weaktemporal}. Since the infinite-noise and infinite-randomness critical behaviors differ qualitatively from each other, novel behavior is expected to emerge if the spatial and temporal disorders are of comparable strength.

\begin{figure}
\includegraphics[width=8cm]{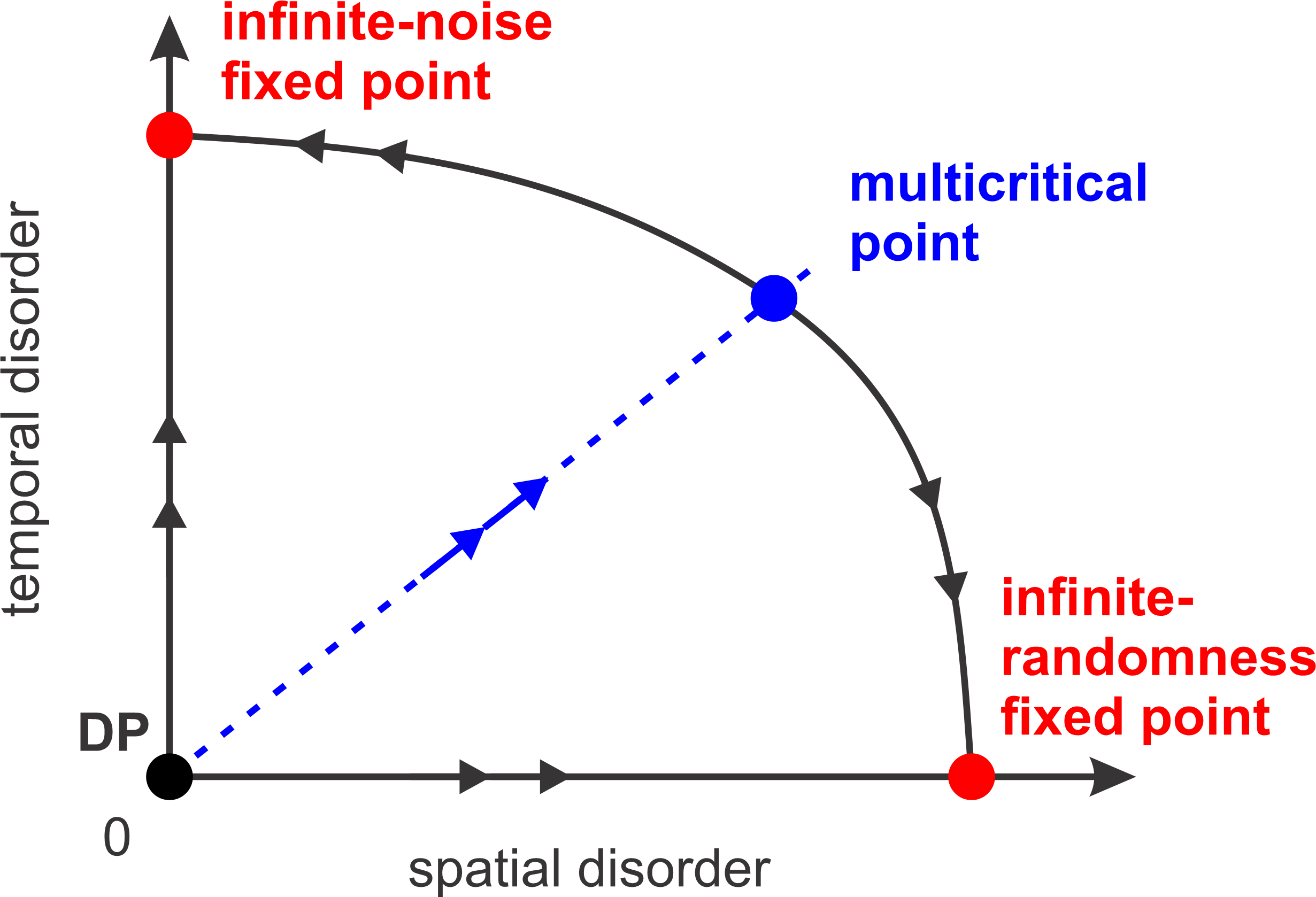}
\caption{Schematic renormalization group flow on the critical manifold spanned by the spatial and temporal disorder strengths.
(DP marks the direct percolation fixed point of the clean contact process.)}
\label{fig:rg_flow}
\end{figure}

The arguably simplest scenario corresponds to the schematic renormalization group flow diagram sketched in Fig.\ \ref{fig:rg_flow}
which contains a multicritical point separating the  infinite-noise and infinite-randomness regimes.
If the ratio of the spatial and temporal disorder strengths
is fine-tuned to be exactly on the separatrix (dashed line) in Fig.\ \ref{fig:rg_flow}, the system flows to the multicritical point under coarse graining. The nonequilibrium phase transition then features novel multicritical behavior. If the system is not exactly on the dashed line, it will eventually flow either to the infinite-noise critical point or to the infinite-randomness critical point. However, if the system is close to (but not exactly on) the dashed line, it will flow towards the multicritical point for a long time before eventually approaching one of the other fixed points. This means the system will show multicritical behavior over a wide transient time interval before eventually crossing over to either infinite-randomness or infinite-noise critical behavior.

Studying the regime where the spatial and temporal disorders are of comparable strength is extremely challenging numerically
because the logarithmically slow dynamics makes it difficult to distinguish the asymptotic behavior from slow crossovers during
the achievable simulation times. In the following, we demonstrate that our numerical data are compatible with the multi-critical
point scenario. We emphasize however, that the unequivocal determination of the fate of the contact process in
this regime is beyond our current computational capabilities.

\begin{figure}
\includegraphics[width=\columnwidth]{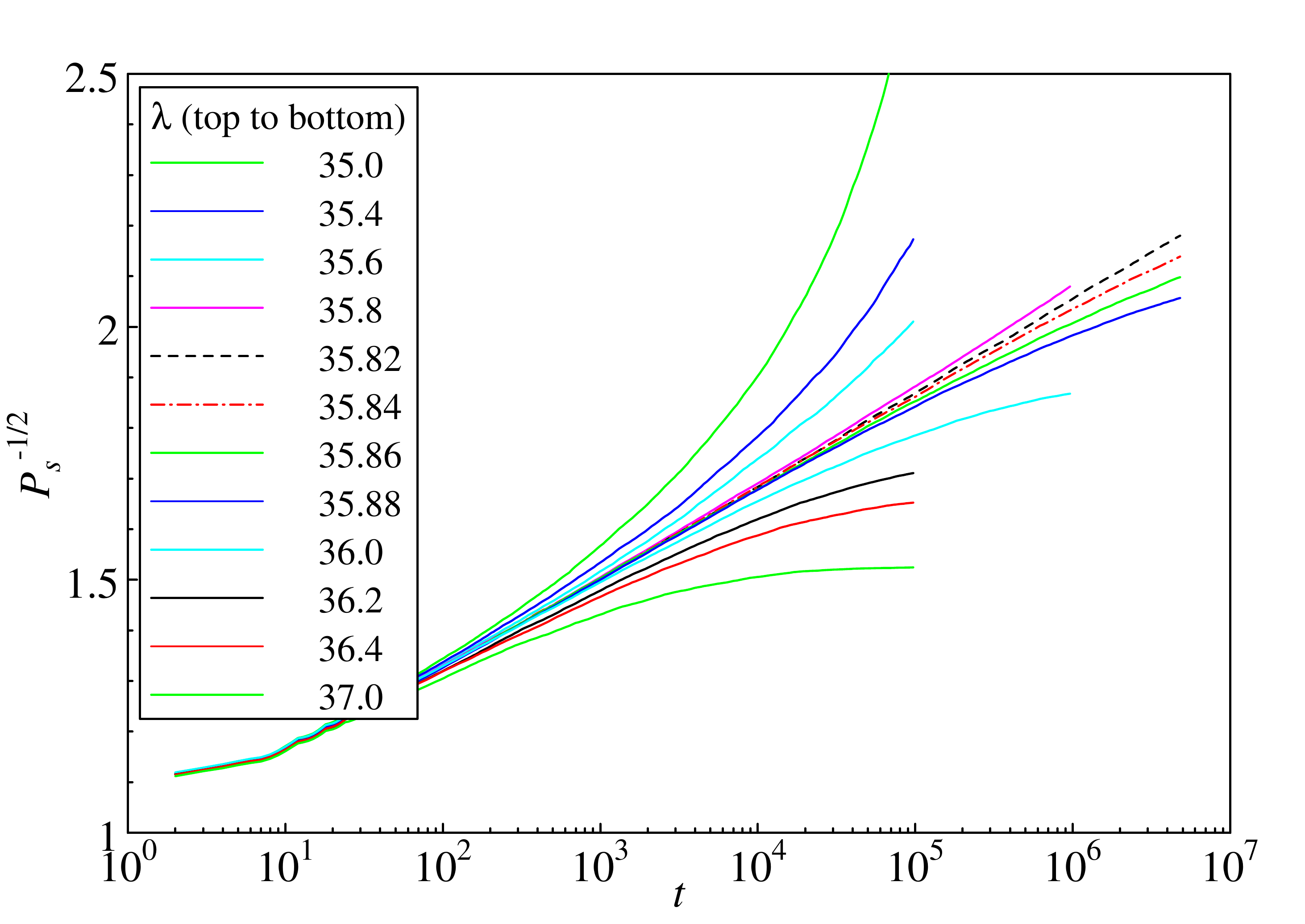}
\caption{Survival probability vs time for $p_t = 0.2$, $c_t=0.05$ $p = 0.2$, $c = 0.05$, and $\Delta t =6$) The data are averages over 20000 disorder configurations with 5 runs per configuration. }
\label{fig:mcp_ps_t}
\end{figure}

\begin{figure}
\includegraphics[width=\columnwidth]{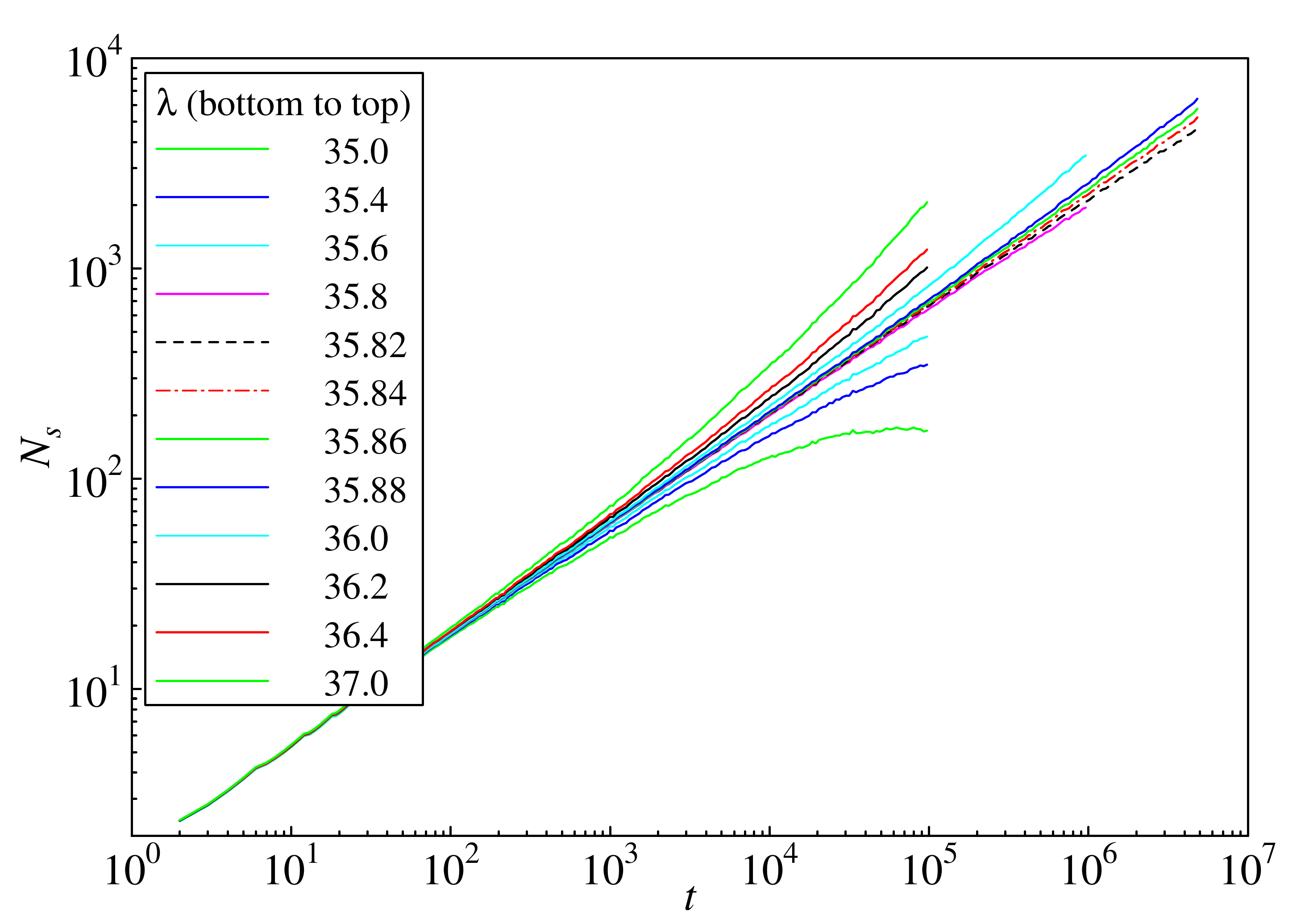}
\caption{Number of active sites vs time for $p_t = 0.2$, $c_t=0.05$ $p = 0.2$, $c = 0.05$, and $\Delta t =6$) The data are averages over 20000 disorder configurations with 5 runs per configuration. }
\label{fig:mcp_ns_t}
\end{figure}

To identify a multicritical system, we start from the sequence of systems studied in Sec.\ \ref{subsec:results-critical-weakspatial} and further increase the spatial disorder by reducing $c$, aiming at identifying a disorder strength for which the (asymptotic) critical behavior differs from both the infinite-randomness and the infinite-noise behavior. As the functional forms of the
observables at the multicritical point are not known, we employ Dickman's \cite{MoreiraDickman96} heuristic criterion of $\lambda_c$
being the smallest $\lambda$ supporting asymptotic growth of $N_s(t)$ to identify the phase transition.
Figures \ref{fig:mcp_ps_t} and \ref{fig:mcp_ns_t} show that the system with $p_t=0.2$, $c_t=0.05$, $p=0.2$, $c=0.05$ approximately fulfills these conditions.
The data at an infection rate of about 35.82 to 35.84 follow the functional forms
\begin{equation}
P_s \sim \ln^{-2}(t)~, \qquad  N_s \sim t^\theta
\label{eq:mcp}
\end{equation}
with $\theta \approx 0.5$ for almost six orders of magnitude in time. These functional forms differ from the behavior in the bulk phases as well as from the infinite-randomness and infinite-noise critical behaviors. This suggests that the parameters $p_t=0.2$, $c_t=0.05$, $p=0.2$,
$c=0.05$ put  the system very close to the separatrix in Fig.\ \ref{fig:rg_flow}, and (\ref{eq:mcp}) approximately represents the multicritical behavior. Small deviations at late times can be attributed to the fact that the system is likely not exactly on the separatrix. To check the consistency of the analysis, we have confirmed that $N_s/R$ behaves as $\ln^{-4}(t)$ as expected from the relation $N_s \sim P_s \rho R^d$.

The multicritical point can also be reached (approximately) by starting from the sequence of systems in Sec.\ \ref{subsec:results-critical-weaktemporal} and further increasing the temporal disorder. The system with
$p_t=0.2$, $c_t=0.12$, $p=0.3$, $c=0.2$ follows the same multicritical behavior (\ref{eq:mcp})  at an infection rate
of $\lambda \approx 16.35$.

The functional forms of Eqs.\ (\ref{eq:mcp}), which combine a logarithmic decay of $P_s$ with a power-law time dependence of $N_s$,
indicate an unconventional type of multicritical point because they appear to be incompatible with the usual scaling laws in which
either $t$ or $\ln t$ appear as scaling variables (but not both). We emphasize, however, that a similar situation already occurs at the infinite-noise critical point of a system with purely temporal disorder, see Eqs.\  (\ref{eq:rho_t_temporal}) and (\ref{eq:Ns_t_temporal}). In that case, the logarithmic time dependence of $P_s$ can be understood as a logarithmic correction to a critical exponent of value zero. A similar scenario may apply to the multicritical point as well.

It is also interesting to note that the decay of $P_s$ at the putative multicritical point, $P_s \sim \ln^{-2}(t)$ is faster than its decay at both the infinite-randomness critical point and the infinite-noise critical point (even though the $P_s$ data are not compatible with an even faster power-law decay). This suggests that when the spatial and temporal disorders are of comparable strength, they weaken each other. The same phenomenon is also observed for the rare region effects and Griffiths singularities discussed in the next section.

\section{Results: Rare regions and Griffiths singularities}
\label{sec:results-griffiths}

In this section, we discuss the effects of rare spatial regions and rare time intervals on the behavior of the contact process
with the combined spatial and temporal disorder of the form $\lambda(x,t) = \lambda_0 f(x) g(t)$.

\subsection{Theory}
\label{subsec:griffiths-combined-theory}

Consider a spatial rare region with an above average $f(x)$. This region can be locally in the active phase even if the bulk system is still inactive. If $f$ follows the binary distribution (\ref{eq:ax}), the strongest rare regions consist of sites with $f=1$ only. As in Sec.\ \ref{subsec:griffiths-spatial}, the probability for finding such a region rare is given by $w(L_{RR}) \sim \exp(- b L_{RR}^d)$ (up to pre-exponential factors). However, the behavior of the lifetime $\tau(L_{RR})$ of such a region depends on the strength of the temporal disorder. If the temporal disorder is sufficiently weak such that the rare region is locally active for all times, $\tau(L_{RR}) \sim \exp(a L_{RR}^d)$ as in the case of purely spatial disorder. For stronger temporal disorder, in contrast,
the rare region will still be mostly active, but inactive during rare time intervals. In this case, the lifetime $\tau(L_{RR})$ depends on $L_{RR}$ via the power law $\tau(L_{RR}) \approx (a L_{RR}^d)^{y}$, as shown in Sec.\ \ref{subsec:griffiths-temporal}.

Inserting $\tau(L_{RR})$ in (\ref{eq:rho_RR}) yields the following
anomalous density decay in the Griffiths phase on the inactive side of the transition:
\begin{eqnarray}
\rho(t) &\sim&  t^{-b/a}=t^{-d/z'}~~~\textrm{weak temporal disorder}~,~  \label{eq:griffithspower2} \\
\rho(t) &\sim&  \exp(-b t^{1/y}/a) ~~~\textrm{strong temporal disorder}~.~  \label{eq:griffithsexp2}
\label{eq:griffiths_rho_both}
\end{eqnarray}
The survival probability $P_s(t)$ in spreading runs behaves in the same manner as $\rho(t)$.
Thus, for sufficiently strong temporal disorder, the power-law Griffiths singularities are weakened and replaced by stretched
exponential behavior. The exponent $y$ is non-universal and depends on how far in the inactive phase a rare region is
during the ''bad'' (low $g(t)$) time periods. $1/y$ is expected to decrease to zero as the transition is approached from the inactive side.

For moderately strong temporal disorder, we expect the Griffiths phase to feature two regions: The
density decay follows a power law for infection rates close to the critical point but a
stretched-exponential behavior for smaller infection rates (i.e., further away from criticality).
With increasing temporal disorder, the power-law region of the Griffiths phase shrinks while the stretched exponential region expands.

Analogous arguments can be made for the Griffiths singularity in the lifetime $\tau_L$ of a finite-size system on the active side
of the phase transition. Consider a system globally in the active phase. Temporal disorder can produce rare time intervals during which the system is temporarily on the inactive side of the transition. For the binary distribution (\ref{eq:bt}), the strongest rare time intervals have $g(t)\equiv c_t$. The probability of finding such time intervals depends exponentially on their lengths, $w(T_{RR}) \sim \exp(-b T_{RR})$, as in Sec.\ \ref{subsec:griffiths-temporal}. However, the time evolution of the density of active sites during these rare time intervals depends on the strength of the spatial disorder.
For weak spatial disorder, the entire system will be in the inactive phase during these intervals, leading to an
exponential density decay, $\rho \sim \exp(-a t)$, as in the case of purely temporal disorder. For stronger spatial disorder,
the system will have spatial regions that remain locally active during the rare time interval, leading to a slower
power-law decay of the density $\rho \sim (at)^{-y}$,

Repeating the analysis of Sec.\ \ref{subsec:griffiths-temporal} for these two cases, we conclude that the
life time $\tau$ of a finite-size system behaves as
\begin{eqnarray}
\tau(L) &\sim&  L^{db/a} = L^{d/\kappa}~ ~~\textrm{weak spatial disorder} ~,~ \label{eq:griffithspower_temporal2}\\
\tau(L) &\sim&  \exp(bL^{d/y}/a) ~ ~ ~\textrm{strong spatial disorder} ~~ \label{eq:griffithsexp_temporal2}
\label{eq:griffiths_tau_both}
\end{eqnarray}
with system size $L$ in the Griffiths phase on the active side of the transition. This means for sufficiently strong spatial disorder, the power-law temporal Griffith singularities of Sec.\ \ref{subsec:griffiths-temporal} are weakened and replaced by
stretched exponentials.

Note that the functional forms (\ref{eq:griffiths_rho_both}) and (\ref{eq:griffiths_tau_both}) have been derived assuming that the relevant rare regions and rare time intervals are uniform in space and time, respectively. This is justified for bounded disorder for which the strongest spatial rare regions have $f(x) \equiv f_{max}$ and the strongest rare time intervals have $g(t) \equiv g_{min}$.
The asymptotic behavior of $\rho$ and $P_s$ for $t \to \infty$ is governed by the strongest rare regions and thus given by
(\ref{eq:griffiths_rho_both}). Along the same lines, the asymptotic behavior of $\tau(L)$ for $L \to \infty$ is governed by the strongest rare time intervals and thus given by (\ref{eq:griffiths_tau_both}). The preasymptotic behavior has contributions from nonuniform rare regions that feature more complicated behavior, leading to nontrivial crossovers.

\subsection{Simulation results}
\label{subsec:griffiths-combined}

\begin{figure}
\includegraphics[width=\columnwidth]{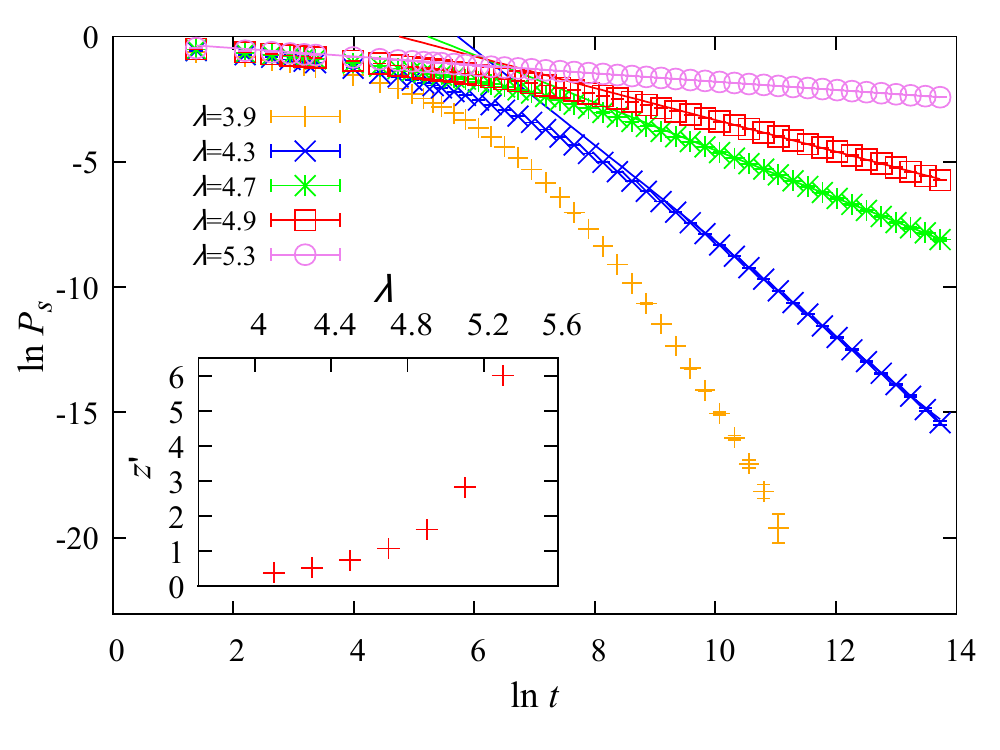}
\caption{Main panel: ln $P_s$ vs ln $t $ for different $\lambda$ below criticality $\lambda_c \approx 5.52$ for $p = 0.3$, $c = 0.2$, $p_t = 0.2$, $c_t = 0.8$ and $\Delta t  =6$. The data are averages over 1000 to 10000 disorder configurations with 100 to $10^5$ runs per configuration. The solid lines are fits to (\ref{eq:griffithspower2}). Inset: Resulting Griffiths exponent $z'$ as a function of the infection rate $\lambda$.}
\label{fig:inactive_power}
\end{figure}

We first consider the survival probability $P_s$ on the inactive side of the transition. To test the power-law Griffiths behavior
(\ref{eq:griffithspower2}), we consider a system with strong spatial disorder but weak temporal disorder
($p = 0.3$, $c = 0.2$, $p_t = 0.2$, $c_t = 0.8$ and $\Delta t=6$).
Figure \ref{fig:inactive_power} presents a double-log plot of $P_s$ vs.\ $t$ for several $\lambda$ below the critical value $\lambda_c \approx 5.52$.
The data indicate that the survival probability follows (\ref{eq:griffithspower2}) for all shown $\lambda \ge 4.3$. Moreover, the Griffiths dynamical exponent $z'$ diverges as the critical infection rate $\lambda_c$ is approached, in agreement with the behavior for purely spatial disorder. For $\lambda=3.9$, in contrast, the data continue to curve downward to the longest times.

These results are in agreement with the scenario discussed in Sec.\ \ref{subsec:griffiths-combined-theory}. To understand this in detail, consider the strongest spatial rare regions which consist of sites with $f\equiv 1$ only. The local infection rate on such a rare region is thus either $\lambda$ or $c_t \lambda =0.8 \times \lambda$. For infections rates $\lambda  > \lambda_c^0/c_t= 4.122$ (where $\lambda_c^0=3.298$ is the clean critical infection rate), the strongest rare regions are thus always on the active side of the clean critical point, explaining the power-law form of the Griffiths singularity.
For $\lambda < \lambda_c^0/c_t$ the rare regions become inactive during the ``bad'' (low $g(t)$) time intervals, putting the system in the stretched-exponential part of the Griffiths phase in which the density decay follows Eq.\ (\ref{eq:griffithsexp2}).

\begin{figure}
\includegraphics[width=\columnwidth]{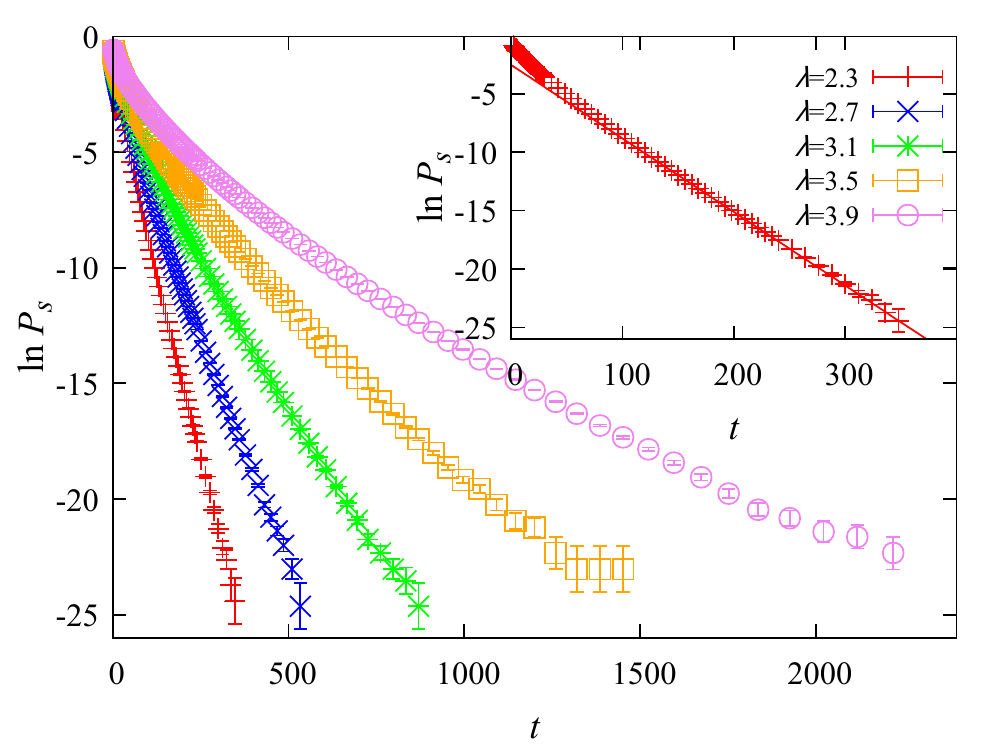}
\caption{Main panel: $\ln P_s$ vs $t$ for different $\lambda$ between 2.3 and 3.9, far from criticality $\lambda_c \approx 7.26$ for $p = 0.3$ and $c = 0.2$, $p_t = 0.2$, $c_t = 0.4$ and $\Delta t =6$). The data are averages over at least than $10^{5}$ disorder configurations with $10^5$ runs each. Inset: Enlarged plot for $\lambda = 2.3$; the linear fit (solid line) confirms a simple exponential decay.}
\label{fig:inactive_small}
\end{figure}

\begin{figure}
\includegraphics[width=\columnwidth]{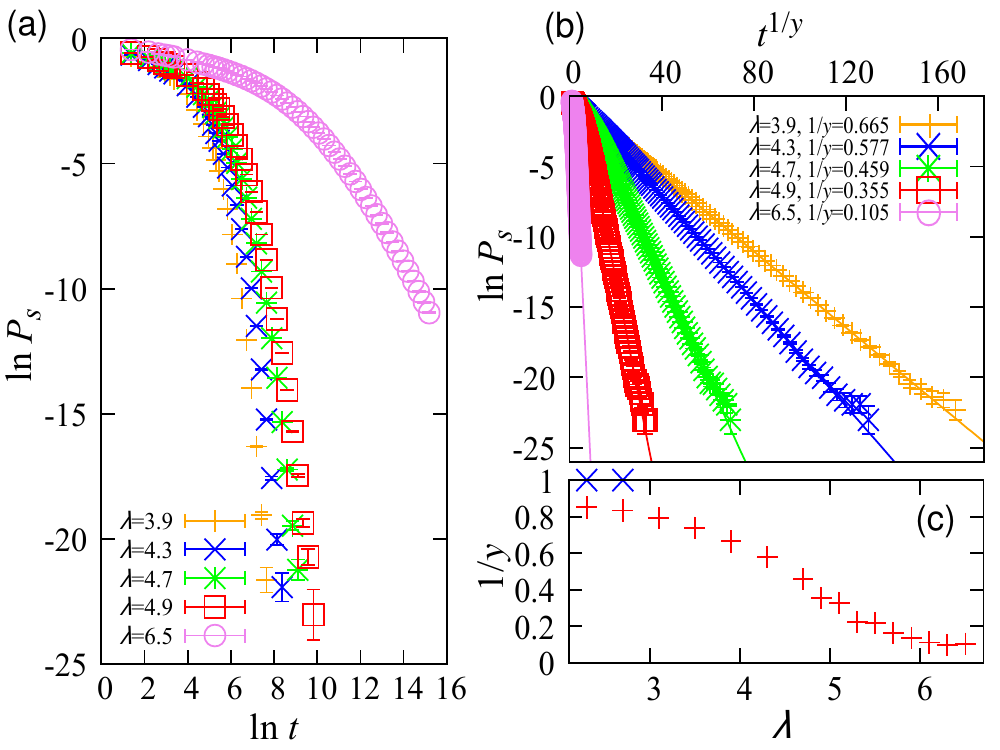}
\caption{(a) $\ln P_s$ vs $\ln t$ for different $\lambda$ below criticality $\lambda_c \approx 7.26$ for $p = 0.3$, $c = 0.2$, $p_t = 0.2$, $c_t = 0.4$ and $\Delta t =6$). The data are averages over $10^4$ to $10^5$ disorder configurations with $10^4$ to $10^5$
runs per configuration. (b) $\ln P_s$ vs  $t^{1/y}$ for the same data. The solid lines are linear fits. (c) Exponent $1/y$ of the stretched exponential (\ref{eq:griffithsexp2}) vs $\lambda$. For $\lambda < \lambda_c^0$, the data can be fitted well with $y=1$, as expected in the conventional inactive phase even though an unrestricted fit yields $1/y$ values slightly below unity.}
\label{fig:inactive_streched_exponent}
\end{figure}

To explore the novel stretched exponential Griffiths behavior (\ref{eq:griffithsexp2}) in more detail, we study a system with stronger temporal disorder, $c_t=0.4$ rather than 0.8. The other parameters remain unchanged ($p = 0.3$, $c = 0.2$, $p_t = 0.2$, and $\Delta t =6$). The critical infection rate for these parameters is  $\lambda_c \approx 7.26$. To cover the entire (inactive) Griffiths phase, we perform simulations for infection rates ranging from 2.3 (below the clean critical value $\lambda_c^0$) to 6.9 close to the phase transition. A semi-log plot of the survival probability for infection rates between 2.3 and 3.9 is shown in Fig.\  \ref{fig:inactive_small}.
For $\lambda$ below the clean critical value $\lambda_c^0=3.298$, the survival probability features a simple exponential decay, as expected in the conventional inactive phase in which there are no locally active rare regions. For $\lambda > \lambda_c^0$, the
system enters the Griffiths phase, and the decay of $P_s$ becomes slower than exponential. However, as is demonstrated via the double-log plot of $P_s$ vs $t$ in Fig.\ \ref{fig:inactive_streched_exponent}(a), the decay for all $\lambda$ in the (inactive) Griffiths phase is faster than a power law.
In fact, all data can be fitted very well with the stretched exponential form (\ref{eq:griffithsexp2}), as shown in Fig.\ \ref{fig:inactive_streched_exponent}(b) which replots the same data in the form $\ln P_s$ vs  $t^{1/y}$ with $y$ chosen such that the data fall onto straight lines. The resulting values of the exponent $1/y$ governing the stretched exponential evolve from unity at the clean critical infection rate $\lambda_c^0$ towards zero at the phase transition. Note that even the strongest rare regions ($f\equiv 1$)
will be inactive during the ``bad'' time intervals everywhere in the Griffiths phase because $c_t \lambda_c < \lambda_c^0$. This explains why the decay of the survival probability takes the stretched exponential form for all infection rates with $\lambda_c^0 < \lambda < \lambda_c$.

\begin{figure}
\includegraphics[width=\columnwidth]{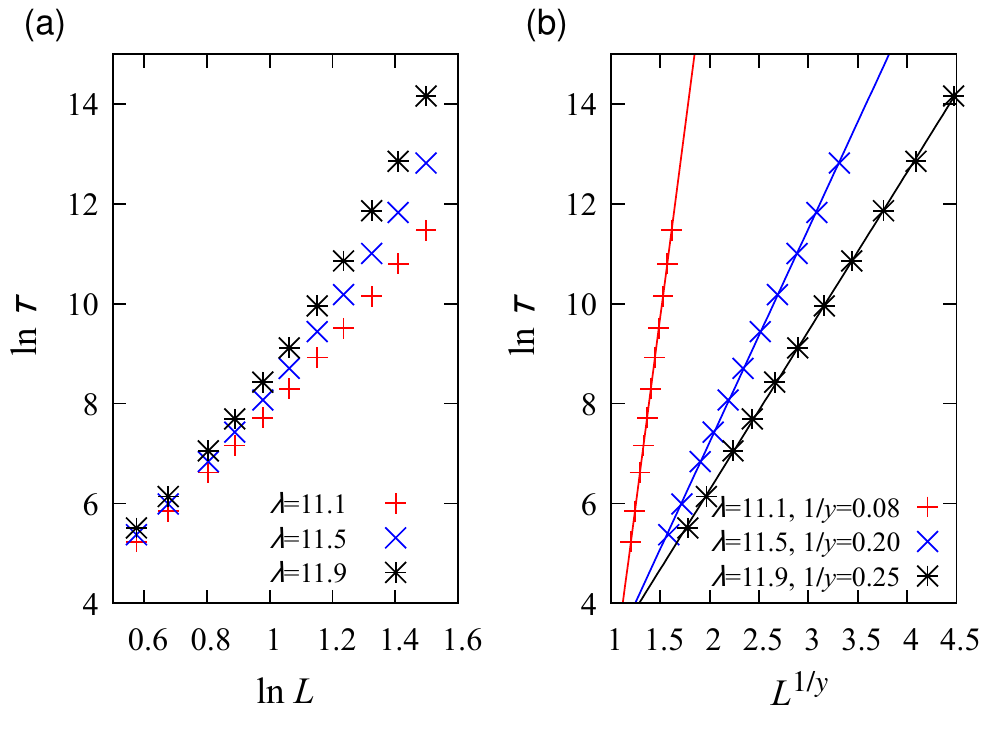}
\caption{(a) Double-log plot of lifetime $\tau$ vs system size $L$ for different $\lambda$ above criticality $\lambda_c \approx 11.08$ for $p = 0.3$, $c = 0.2$, $p_t = 0.2$, $c_t = 0.2$ and $\Delta t =6$. The data are determined from decay runs, averaged over 10240 disorder configurations (one run per configuration). (b) The same data plotted as $\ln \tau$ vs $L^{1/y}$, with $y$ chosen such that the data fall onto straight lines.}
\label{fig:active_lifetime}
\end{figure}

We now turn to the behavior of the lifetime of a finite-size system on the active side of the transition. The goal is to
test wether the power-law temporal Griffiths behavior (\ref{eq:griffithspower_temporal2}) gets replaced by the stretched exponential
(\ref{eq:griffithsexp_temporal2}) if sufficiently strong spatial disorder is added to the temporally disordered contact process.
Figure \ref{fig:active_lifetime}(a) shows a double log plot of the lifetime vs system size for $p = 0.3$, $c = 0.2$, $p_t = 0.2$, $c_t = 0.2$, and $\Delta t=6$ at infection rates slightly above the critical value  $\lambda_c \approx 11.08$.
The figure demonstrates that the increase is faster than a power law. The same data are replotted in Fig.\ \ref{fig:active_lifetime}(b) in the form $\ln \tau$ vs $L^{1/y}$ motivated by Eq.\ (\ref{eq:griffithsexp_temporal2}). For properly chosen $y$-values, all data fall onto straight lines, confirming that the lifetime follows the stretched exponential Griffiths behavior (\ref{eq:griffithsexp_temporal2}). The exponent $1/y$ increases with increasing distance from criticality, as expected.

\section{Conclusions}
\label{sec:conclusions}

In summary, we have investigated the combined influence of spatial and temporal random disorder on the absorbing-state
phase transition in the one-dimensional contact process. Specifically, we have studied the case of decoupled spatial
and temporal disorders for which the local infection rates $\lambda(x,t)$ are the product of a purely spatial term and
a purely temporal term, $\lambda(x,t) = \lambda_0 f(x) g(t)$. In contrast to completely uncorrelated spatiotemporal
randomness, such disorder which contains infinite-range correlations in space and time
is a relevant perturbation at the clean DP critical point.

We have employed a generalization of the Harris criterion \cite{VojtaDickman16} to predict that
the infinite-randomness critical point of the spatially disordered contact process is stable against weak
temporal disorder. Analogously, the criterion predicts that the infinite-noise critical point of the temporally
disorder contact process is stable against weak spatial disorder. We have confirmed these predictions by
extensive computer simulations. In the interesting parameter region where both disorders are of comparable strength,
the critical behavior appears to differ from both the infinite-randomness and infinite-noise critical behaviors.
Our simulation data are compatible with the simplest scenario in which a single multicritical point separates
the infinite-randomness and infinite-noise regimes. However, due to the very slow dynamics of the contact process
in the presence of both disorders, we cannot exclude more complicated scenarios that involve novel critical behavior
in an extended parameter region. In the absence of theoretical predictions, the complete quantitative understanding
of the (multi)critical behavior from simulations would require simulation times several orders of
magnitude larger than what is achievable today. This problem thus remains a task for the future.

In addition to the nonequilibrium phase transition itself, we have also investigated the effects of rare regions and
rare time intervals in the Griffiths phases near the transition. By means of optimal fluctuation arguments, we have shown
that adding weak temporal disorder does not change the power-law Griffiths behavior of the density and survival probability
of the spatially disordered contact process on the inactive side of the transition (at least sufficiently close to the transition). Stronger temporal disorder, in contrast, weakens the ``spatial'' Griffiths singularity in the density and
survival probability, replacing the slow power-law decay with a faster stretched exponential. The behavior of the lifetime
as a function of system size in the ``temporal'' Griffiths phase on the active side of the transition is completely analogous.
Adding weak spatial disorder to the temporally disordered contact process does not change the power-law Griffiths
behavior but sufficiently strong spatial disorder weakens the singularity from power-law to stretched exponential behavior.
The notion that the spatial and temporal disorders weaken each other is also consistent with the observation that the decay
of the survival probability with time at the putative multicritical point is faster than the decay at either the
infinite-randomness critical point or the infinite-noise critical point.

Our explicit computer simulation results are for one space dimension. However, the stability arguments based on
the generalized Harris criterion apply equally to one, two, and three space dimensions.
The same applies to the optimal fluctuation
arguments governing the Griffiths singularities. We therefore expect most of our qualitative results to carry over
from one  to two and three space dimensions.

Recently, Odor \cite{Odor99} studied the stability against temporal disorder
of the Griffiths phase in a threshold model running on a large
human connectome graph. As in our problem, he found that the (spatial) Griffiths phase is
insensitive to weak temporal disorder while sufficiently strong temporal disorder suppresses the power-law Griffiths
singularities.

Clearcut experimental examples of absorbing-state transitions were missing for a long time \cite{Hinrichsen00b}.
By now, such transitions have been observed,  however,  in turbulent liquid crystals \cite{TKCS07},
driven suspensions \cite{CCGP08,FFGP11}, growing bacteria colonies \cite{KorolevNelson11,KXNF11}, and in
the dynamics of superconducting vortices \cite{OkumaTsugawaMotohashi11}. Studying these systems under the
combined influence of spatial disorder and external noise will permit experimental tests of our results.
The influence of environmental fluctuations and inhomogeneities on the extinction of a biological population
are attracting considerable attention today in the contexts of both epidemic spreading and of global warming
and other large-scale environmental changes (see, e.g., Ref.\ \cite{OvaskainenMeerson10}).
In the laboratory, these questions could be analyzed, e.g., by growing bacteria or yeast populations in
spatially inhomogeneous environments and fluctuating external conditions.

\acknowledgments

This work has been supported in part by the National Science Foundation under Grant Nos.\  DMR-1828489 and
OAC-1919789. The simulations were performed on the Pegasus and Foundry clusters at Missouri S\&T.

\bibliographystyle{apsrev4-2}
\bibliography{rareregions}

\end{document}